\documentclass[acmsmall,dvipsnames]{acmart}

\usepackage{algorithm}
\usepackage[binary-units]{siunitx}

\newcommand{\scamac}{\textsf{ScaMaC}}
\newcommand{\ghost}{\textsf{GHOST}}
\newcommand{\mpi}{\textsf{MPI}}
\newcommand{\mpix}{\textsf{MPI+X}}
\newcommand{\openmp}{\textsf{OpenMP}}
\newcommand{\blas}{\textsf{BLAS}}

\acmJournal{TOPC}

\begin{document}



\author{Andreas Alvermann}
\orcid{0000-0002-0367-2721}
\affiliation{\department{Institut f\"ur Physik}\institution{Universit\"at Greifswald}\streetaddress{Felix-Hausdorff-Stra{\ss}e 6}
\city{17489 Greifswald}\country{Germany}}
\email{alvermann@physik.uni-greifswald.de}
\authornote{Author to whom any correspondence should be addressed.}
\author{Georg Hager}
\orcid{0000-0002-8723-2781}
\affiliation{\department{Erlangen National High Performance Computing Center}\institution{Friedrich-Alexander-Universit\"at Erlangen-N\"urnberg}\streetaddress{Martensstra{\ss}e 1}
\city{91058 Erlangen}\country{Germany}}
\email{georg.hager@fau.de}
\author{Holger Fehske}
\orcid{0000-0003-2146-8203}
\affiliation{\department{Institut f\"ur Physik}\institution{Universit\"at Greifswald}\streetaddress{Felix-Hausdorff-Stra{\ss}e 6}
\city{17489 Greifswald}\country{Germany}}
\affiliation{\department{Erlangen National High Performance Computing Center}\institution{Friedrich-Alexander-Universit\"at Erlangen-N\"urnberg}\streetaddress{Martensstra{\ss}e 1}
\city{91058 Erlangen}\country{Germany}}
\email{fehske@physik.uni-greifswald.de}

\title{Orthogonal layers of parallelism  in large-scale eigenvalue computations}
\begin{abstract}
  We address the communication overhead of distributed sparse 
  matrix-(multiple)-vector multiplication in the context of large-scale
  eigensolvers, using filter diagonalization as an example.  The basis
  of our study is a performance model which includes a communication
  metric that is computed directly from the matrix sparsity pattern
  without running any code.  The performance model quantifies to which
  extent scalability and parallel efficiency are lost due to communication
  overhead.

  To restore scalability, we
  identify two orthogonal layers of parallelism in the filter
  diagonalization technique.  In the horizontal layer the
  rows of the sparse matrix are distributed across individual processes. In the
  vertical layer bundles of multiple vectors are distributed across separate process groups.
  An analysis in terms of the communication metric predicts that
  scalability can be restored if, and only if, one implements the two
  orthogonal layers of parallelism via different distributed vector
  layouts.

  Our theoretical analysis is corroborated by benchmarks for
  application matrices from quantum and solid state physics,
  road networks, and nonlinear programming.
  We finally demonstrate the benefits of using orthogonal layers of
  parallelism with two exemplary application cases---an exciton and a
  strongly correlated electron system---which incur either small or
  large communication overhead.
\end{abstract}
\maketitle

\section{Introduction}

Research into complex systems often involves large-scale eigenvalue problems defined by sparse symmetric (or Hermitian) matrices.
A large number of algorithms have been developed to solve such eigenproblems~\cite{So02,Saad11,ARPACK,Hernandez:2005:SSF,Anasazi}.
While the computation of a few extremal eigenvalues is usually a manageable task, the extraction of many interior or nearly-degenerate eigenvalues required in some applications remains a computational challenge~\cite{SBR08,10.21468/SciPostPhys.5.5.045,PhysRevLett.125.156601,10.21468/SciPostPhys.11.2.021}.
In this context, techniques such as subspace iteration or filter diagonalization (FD) have resurfaced~\cite{ZSTC06,Pol09,Pieper16,10.1145/3313828,HuberNLA} as an alternative to the Lanczos or Jacobi-Davidson algorithms~\cite{CW85,SV96,FSV98,zoellner2015}.

Efficient, cost-effective computations require good algorithms and
performant implementations. For the parallelized solution of large-scale eigenvalue problems the main issues are the performance and the parallel efficiency of the two central operations, sparse matrix-vector multiplication (SpMV) and orthogonalization.
Beyond the ubiquitous memory bandwidth bottleneck, the main limiting factor is communication among processes.
While communication-avoiding techniques exist for orthogonalization~\cite{SW02,DGHL12}
and for matrix power kernels~\cite{10.1145/1654059.1654096,AlappatMPK2023},
communication cannot be avoided in distributed SpMV.

\begin{figure}
\hspace*{\fill}
\fbox{\includegraphics[width=0.25\textwidth]{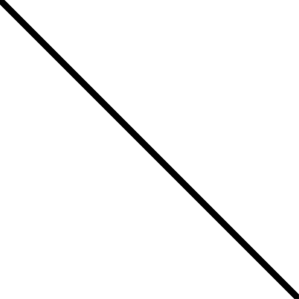}}
\hspace*{\fill}
\fbox{\includegraphics[width=0.25\textwidth]{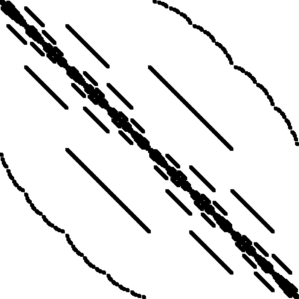}}
\hspace*{\fill}
\caption{Sparsity pattern of the \texttt{Exciton200} (left panel, dimension $D = 193\,443\,603$) and \texttt{Hubbard16} (right panel, dimension 
$D = 165\,636\,900$) matrices from the scalable matrix collection \scamac{}~\cite{SCAMAC}.
}
\label{fig:sparsity}
\end{figure}

The significance of communication, mainly quantified by the communication volume per SpMV, depends on the sparsity pattern of the matrix.
Two typical matrices from quantum physics applications~\cite{Da94,StronglyCorrelated,ALF11,ExcitonII} are shown in Fig.~\ref{fig:sparsity}.
Note that for realistic applications in quantum physics, the matrices have dimensions in the hundreds of millions.
For the ``narrow'' sparsity pattern of the \texttt{Exciton200} matrix we expect that communication is insignificant and can overlap with computation and local memory access.
For the ``wide'' sparsity pattern of the \texttt{Hubbard16} matrix we expect that communication dominates the SpMV to such an extent that scalability and parallel efficiency are lost.
We will later be able to differentiate these cases with a communication metric (see Sec.~\ref{sec:metric}).

In the present paper we describe how scalability and parallel efficiency can be restored if one recognizes that an eigensolver such as FD, and in fact any block eigensolver with multiple search vectors, possesses two orthogonal layers of parallelism:
a ``horizontal layer,'' where each vector is distributed across multiple processes,
and a ``vertical layer,'' where bundles of search vectors are distributed across different process groups.
Note that the eigenvector search space typically comprises several tens or hundreds of vectors.
The two layers allow us to choose and compare different layouts of the distributed vectors.
This new option can result in a significant reduction of communication in SpMV (and therefore better scalability), although at the price of increased memory consumption if the matrix has to be stored redundantly.

Making use of the two layers of parallelism is not an entirely trivial task since SpMV and orthogonalization have competing communication patterns: Distributed vector layouts that are optimal for SpMV do not coincide with those optimal for orthogonalization. 
The first objective of the present paper is, therefore, to introduce communication metrics that quantify the trade-off between communication during SpMV and during redistribution of vectors between different layouts.
The second objective is to validate our communication metrics with extensive benchmarks for real application matrices, such as those depicted in Fig.~\ref{fig:sparsity}. The third objective is to study the performance benefits of using the two layers of parallelism in large-scale eigenvalue computations, here with the FD scheme.

Note that the present paper has no intention to address algorithmic issues of FD\@.
We use FD as an example of an eigensolver that allows us to reliably compute several or many eigenvalues even for complicated matrices
and to apply our analysis of communication in SpMV in a relevant computational context.
While FD provides the present computational context, our performance analysis in terms of the communication metrics is not restricted to FD but extends to any block eigensolver where SpMV applied to multiple vectors is a dominant operation. 
Therefore, we restrict ourselves to the plain FD scheme, even though we are aware that it will not serve as a general-purpose eigensolver, and postpone algorithmic considerations to future work.

The paper is organized from this viewpoint.
In Sec.~\ref{sec:FD} we recall the basics of FD in the context of large-scale computations.
In Sec.~\ref{sec:layers} we describe the two orthogonal layers of parallelism and introduce the communication metrics that are pivotal to our performance analysis.
We use the communication metrics to analyze a number of SpMV benchmarks. Then, we discuss how different vector layouts allow us to restore scalability and parallel efficiency of SpMV. In Sec.~\ref{sec:FDlayers} we validate this approach using FD for the solution of large-scale eigenvalue problems in the exciton and Hubbard application cases.
Finally, we summarize in Sec.~\ref{sec:summary}.
The appendix provides further benchmarks for various other application matrices.

\section{Outline of filter diagonalization}
\label{sec:FD}

\begin{figure}
\hfill
\includegraphics[scale=0.8]{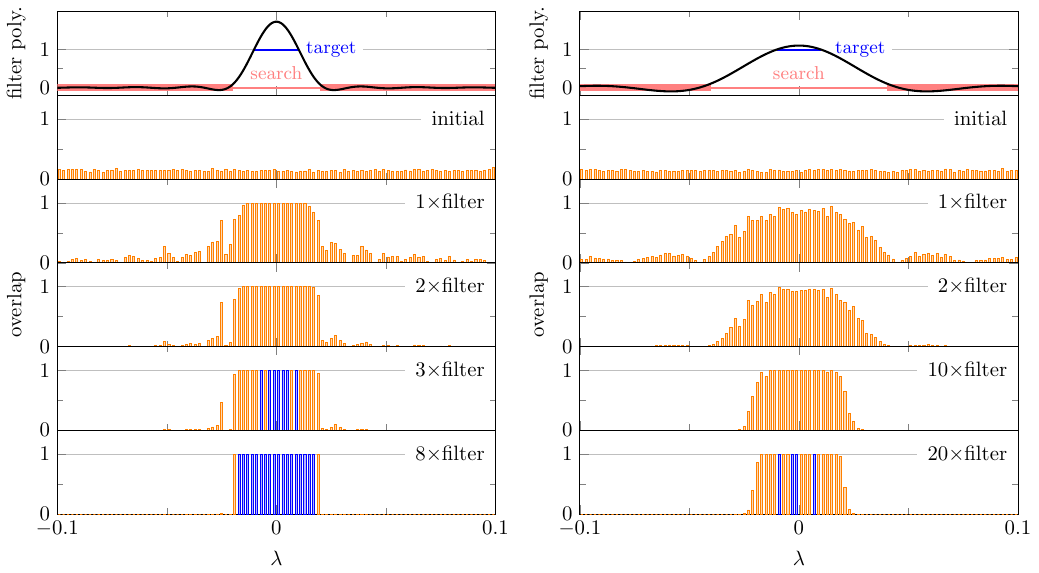}
\hfill{}
\caption{Filter diagonalization for $1000$ equidistant eigenvalues in the interval $[-1,1]$.
The top panels show the filter polynomial for target interval $[-0.01,0.01]$ (with $N_t=10$) and search interval $[-0.02,0.02]$ (left column) or $[-0.04,0.04]$ (right column). 
Outside of the search interval the filter polynomial is small, as indicated by the red boxes which contain the polynomial. The lower panels show the overlap of the search space (with $N_s=20$ vectors) with the true eigenvectors. In the left column, where the search interval contains fewer than $N_s$ eigenvalues, filter diagonalization converges with eight applications of the polynomial filter (converged vectors with overlap above $1-\epsilon$, for $\epsilon=10^{-7}$, are shown in blue; converged eigenvalues cover the target interval and may extend into the search interval).
In the right column, where the search interval contains more than $N_s$ eigenvalues, filter diagonalization does not converge.
}
\label{fig:applyfilter}
\end{figure}

FD proceeds by repeated application of a (polynomial) filter to a search space of $N_s$ independent vectors.
Each application of the filter amplifies (suppresses) targeted (unwanted) eigenvectors of a given matrix $A$, until a subset of $N_t$ vectors has converged to the eigenvectors corresponding to the eigenvalues closest to a target $\tau$.
This scheme is illustrated in Fig.~\ref{fig:applyfilter}.

For large-scale problems, where matrix inversion or factorization~\cite{PARDISO, ILUPACK} is not an option, the restriction to polynomial filters is unavoidable.
A polynomial filter maps search vectors $v \mapsto p[A] \, v$,
where the filter polynomial $p(\cdot)$ is constructed to be large (small) inside (outside) of a target interval around the target $\tau$. 
Since $ p[A]$ is a polynomial in the matrix $A$ it can be evaluated with SpMV\@.

FD works equally for extremal and interior eigenvalues, although convergence for extremal eigenvalues is considerably faster than for interior eigenvalues~\cite{Saad11,Pieper16}.
The large-scale computations considered here, where the target interval is a small fraction of the entire spectrum, easily require a polynomial degree of order $10^3$ or higher.

Convergence of FD depends not only on the quality of the filter polynomial but also on the number of search vectors $N_s$. For fast convergence one should choose $N_s \gg N_t$. Often, $N_s / N_t \simeq 4$ is a reasonable choice (see~\cite{Pieper16} for an extended analysis).
Therefore, for a $D$-dimensional eigenproblem, the effective problem size in FD is $D \times N_t$ if $N_t \ge 1$ eigenvectors are requested, and the effective computational size is $D \times N_s$ instead of just $D$.
As discussed in Sec.~\ref{sec:layers}, this allows us to parallelize the problem both along the ``vertical'' axis $D$ and the ``horizontal'' axis $N_s$, which results in two orthogonal layers of parallelism.

Note that we do not split the problem into multiple independent (``embarrassingly parallel'') computations with a smaller number of target and search vectors, as would be done in a ``spectrum slicing'' approach~\cite{doi:10.1137/18M1170935}. An essential lesson 
from the analysis in, e.g., Ref.~\cite{Pieper16} 
is that a reduction of the size of the search space is paid for with a disproportionate increase in the degree of the filter polynomial, and hence of the total computational effort.
Convergence of FD for large-scale problems calls for a large search space.

\begin{algorithm}
\caption{Outline of a minimal FD algorithm, with redistribution of vectors between the stack layout and the panel layout in steps~7,~9.
}
\label{alg:peigfex}
\includegraphics[scale=0.85]{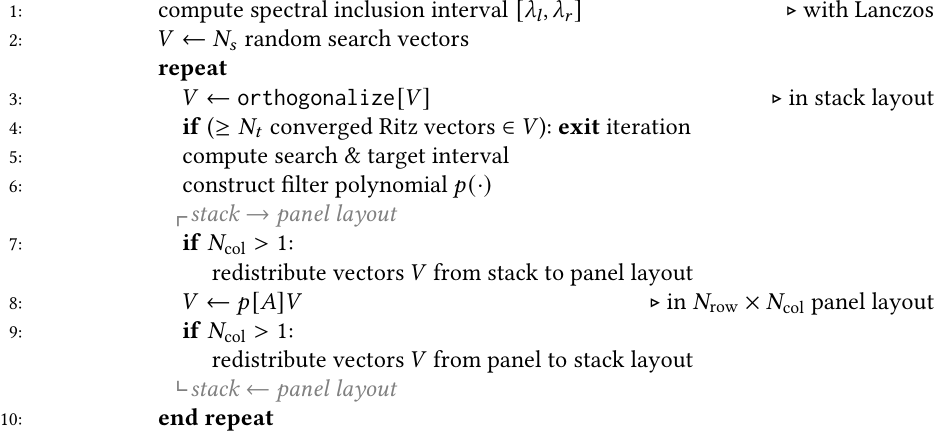}
\end{algorithm}

Starting from the illustration of the FD scheme in Fig.~\ref{fig:applyfilter} we arrive at Algorithm~\ref{alg:peigfex}.
The algorithm already contains steps 7 and 9 associated with the two layers of parallelism that will be discussed in Sec.~\ref{sec:layers}.
Apart from these steps, the algorithm comprises only standard operations.

The preparatory phase (steps 1,\,2) computes (using, e.g., a few steps of the Lanczos algorithm) a spectral inclusion interval $[\lambda_\mathrm{l},\lambda_\mathrm{r}]$ that is required for the polynomial filter in Chebyshev form~\cite{WWAF06}.
The FD iteration alternates between orthogonalization (step 3) and the polynomial filter (step 8).
The algorithm terminates successfully when $N_t$ eigenvectors with eigenvalues in the target interval have converged.
To detect this, we measure the residual $\rho = \| A \vec v- \lambda \vec v \|_2$ of the current best $N_t$ approximate Ritz vectors in the search space and demand $\rho \le 10^{-10}$ for convergence.

During the FD iteration the algorithm has to keep track of the target and search intervals $[\tau_\mathrm{l},\tau_\mathrm{r}]$  and $[\sigma_\mathrm{l},\sigma_\mathrm{r}]$, which are not known in advance.
The target (search) interval must contain at least $N_t$ (at most $N_s$) eigenvalues of $A$.
If the search interval is too large, FD fails to converge, as shown in the right column of Fig.~\ref{fig:applyfilter};
if it is too small, the required degree of the filter polynomial becomes unnecessarily high.
Different strategies for the selection of the target and search intervals have a profound impact on the total number of FD iterations, which leaves much room for algorithmic optimization.
Currently, we determine both intervals from Lehmann intervals~\cite{Lehmann63}.

For a given target and search interval, the filter polynomial is constructed from the Chebyshev expansion $p(x) = \sum_{k=0}^n \mu_k T_k(x)$ of a window function (see Ref.~\cite{Pieper16} for definitions), or from a more general polynomial approximation~\cite{improvedcoeff,doi:10.1137/060648945}.
The evaluation of the polynomial filter $V \mapsto p[A] V$ uses the Chebyshev iteration in Algorithm~\ref{alg:chebfilter}.
The polynomial filter is applied simultaneously to all search vectors, which are given as the columns of a $D \times N_s$ matrix $V$.
For polynomial degree $n$, the evaluation requires $n$ SpMVs and \blas{} level 1 vector-vector operations.

Note that the SpMVs in Algorithm~\ref{alg:chebfilter} are executed simultaneously for multiple vectors $V$ in the form of a ``sparse matrix-multiple-vector multiplication,'' which allows for extensive node-level performance engineering through the reduction of memory traffic~\cite{10.1007/978-3-319-92040-5_17}.
Conversely, this implies that
the communication overhead of distributed SpMV is the single limiting factor for scalability and parallel efficiency of the polynomial filter.

After the initial assignment of random vectors and after each application of the polynomial filter, the search vectors need to be orthogonalized. Afterwards, a single additional SpMV provides the Ritz values and Lehmann intervals used to check for convergence and to determine the current target and search interval.
For orthogonalization, a communication-avoiding technique such as the SVQB method~\cite{SW02} or the tall skinny QR decomposition (TSQR)~\cite{DGHL12} should be used.
We use TSQR because of its superior numerical stability for a large number of search vectors.

\begin{algorithm}
\caption{Evaluation of the polynomial filter $V \mapsto p[A] V$ for a filter polynomial $p(x) = \sum_{k=0}^n \mu_k T_k(x)$, given as a Chebyshev expansion of degree $n \ge 2$. The algorithm requires two matrices $W_1, W_2$ with the same size as $V$ as workspace for the Chebyshev iteration. The scaling parameters $\alpha, \beta$ are determined from the spectral inclusion interval $[\lambda_l, \lambda_r]$.
The SpMV and \texttt{axpy} operations in step 7 are fused into a single compute kernel to avoid reloading $W_2$ from memory~\cite{10.1007/978-3-319-92040-5_17}.
}
\label{alg:chebfilter}
\includegraphics[scale=0.85]{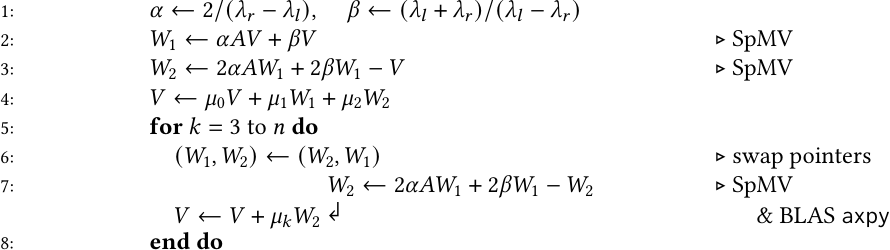}
\end{algorithm}

\section{The two orthogonal layers of parallelism}
\label{sec:layers}

The FD iteration alternates between two operations with competing communication patterns, the polynomial filter and orthogonalization.
According to the discussion in Sec.~\ref{sec:FD}, evaluation of the polynomial filter requires many SpMVs ($\propto n$) for many vectors ($\propto N_s$).
For large problem sizes (vector dimension $D \gtrsim 10^8$, number of search vectors $N_s > N_t \gtrsim 10$),
FD cannot be executed on a single compute node with limited memory resources but requires distributed computing.
This is the setting for the performance analysis in the present section.

\begin{figure}
\hfill
\includegraphics[scale=0.9]{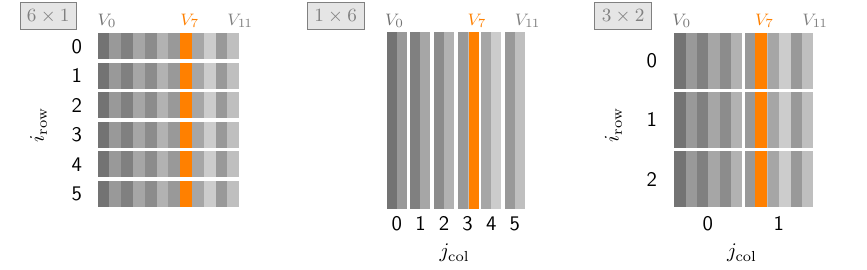}
\hfill{}
\caption{Illustration of the (from left to right) $6 \times 1$ panel = ``stack'', $1 \times 6$ panel = ``pillar'', and $3 \times 2$ panel layout for $N_s = 12$ column vectors $V_0, \dots, V_{11}$ distributed across $\mathcal P=6$ processes.
Orthogonalization (or SpMV) requires 
communication of vector entries along the horizontal (or vertical) axis.
This leads to two orthogonal layers of parallelism in the FD algorithm.}
\label{fig:layouts}
\end{figure}

The key decision in distributed sparse linear algebra is how to distribute vectors and matrices
across the parallel processes. 
Note that by ``process'' we will usually mean an \mpi{} process in a hybrid \mpix{} program (often, \textsf{X}=\openmp{}).
One \mpi{} process may span an entire compute node or is associated with an individual non-uniform memory access (NUMA) locality domain.
The basic assumption is that communication between different processes crosses node or NUMA boundaries and is significantly slower than shared-memory access within a single process.

Three different distributed vector layouts are depicted in Fig.~\ref{fig:layouts}, where the FD search vectors are assembled as the columns of a $D \times N_s$ matrix $V = (V_0, V_1, \dots, V_{N_s-1})$.
These vectors have to be distributed across $\mathcal P$ processes, where $\mathcal P$ denotes the total number of processes at our disposal.

In the standard ``stack'' layout as depicted in the left panel in Fig.~\ref{fig:layouts}, the $N_s$ search vectors are sliced horizontally such that each of the $\mathcal P$ processes keeps a fraction $1/\mathcal P$ of every search vector, i.e., $D/\mathcal P$ rows of the matrix $V$.
In this layout, orthogonalization requires minimal communication. For example,
computation of the Gram matrix $V^t V$ needed in SVQB orthogonalization~\cite{SW02}
requires only communication of local scalar products in a global \verb.MPI_Allreduce. operation.
The aggregated communication volume is of order $\mathcal P \times N_s^2$ and independent of $D$.
The TSQR scheme is more involved but still requires only minimal communication~\cite{DGHL12}.
SpMV, on the other hand, needs to communicate the vector entries between processes.
The communication volume depends on the sparsity pattern of the matrix at hand.
Already visual inspection of Fig.~\ref{fig:sparsity} reveals that the communication volume of SpMV with the \texttt{Hubbard} matrix exceeds that with the stencil-like \texttt{Exciton} matrix.
The difference is quantified by the $\chi$ communication metrics introduced below.

In the alternative ``pillar'' layout shown in the central panel of Fig.~\ref{fig:layouts}, each process keeps a bundle of $N_s/\mathcal P$ entire search vectors. Pictorially, we slice the matrix $V$ of search vectors vertically instead of horizontally.
In this layout, SpMV does not require any inter-process communication.
Orthogonalization, however, now entails communication of entire vectors (e.g., to compute a scalar product).
The minimal communication volume is of the order $D \times N_s \times (1-1/\mathcal P)$ vector entries, e.g., to assemble all vectors at one process, which likely exceeds that of SpMV in the stack layout.

The stack (with $N_\mathrm{col}=1$) and pillar (with $N_\mathrm{row}=1$) layout are special cases of the general ``panel'' vector layout depicted in the right panel of Fig.~\ref{fig:layouts}.
For the panel layout,
we arrange the $\mathcal P$ processes as a rectangular 
$N_\mathrm{row} \times N_\mathrm{col}$ Cartesian grid.
We can
freely vary the number of process rows and columns to obtain differently shaped 
panel layouts,
provided that the total number of processes $\mathcal P = N_\mathrm{row} \times N_\mathrm{col}$, which depends on the available compute resources, remains fixed.

 Distribution of vectors according to the Cartesian grid of the panel layout introduces the two orthogonal (horizontal vs.\ vertical) layers of parallelism.
Since orthogonalization requires communication along the horizontal axis and SpMV along the vertical axis of Fig.~\ref{fig:layouts}, the panel layout can achieve a trade-off between the competing communication patterns of orthogonalization and the polynomial filter in the FD iteration.
Whether a trade-off is achieved and what speedup of the whole FD iteration can be expected with a panel layout depends on a number of parameters, including the $\chi$ communication metric for SpMV introduced below.

Together with the vectors the sparse matrix has to be distributed among the processes, since the matrix elements are required locally where the computations are performed. Therefore, the vector layout determines the layout of the sparse matrix: Each process has to store the sparse matrix elements for the respective rows of the dense matrix of vectors. As can be seen in Fig.~\ref{fig:layouts}, all rows are represented in each process column.
Therefore, also the sparse matrix has to be duplicated in each process column.
In the stack layout, a single copy of the sparse matrix is distributed among all processes. In the pillar layout, each process keeps a full copy of the sparse matrix.
The duplication of the matrix is not a weakness of the specific vector layouts chosen here; it cannot be avoided if communication \emph{within} the SpMV operation should be reduced.
In essence, one has to trade memory for performance---unless, of course, a matrix-free SpMV implementation is available for the problem at hand. 

Note that a panel layout of vectors or matrices is used in other contexts, e.g., in the block-cyclic data layout of \textsf{ScaLAPACK}~\cite{SCALAPACK}.
In difference to \textsf{ScaLAPACK}, we here deal with sparse instead of dense linear algebra operations.
In particular, although the vector layout prescribes how the sparse matrix has to be distributed, the vector and matrix layouts do not coincide.
The sparse matrix is always distributed rowwise and never partitioned columnwise. Any partition of the matrix columns would result in additional communication of matrix elements, simply because they would not be available locally where the computations are performed, and thus invalidate all efforts to reduce communication.
To reflect these differences, we introduced the specific terminology of the ``stack,'' ``panel,'' and ``pillar'' layouts. They correspond to a 1D rowwise, 2D, and 1D columnwise partitioning for the vectors, but not for the sparse matrix.
Furthermore, depending on the sparsity pattern, SpMV results in irregular communication patterns that need to be captured by appropriate metrics yet to be defined.
We also have to consider switching between different vector layouts to accommodate the competing communication patterns of SpMV and orthogonalization.

The purpose of the present section is to establish the three performance-oriented messages of the paper, which are independent of the application-centered aspects addressed in the next section with the example of the FD algorithm. The three messages are (i) there are metrics that allow for the a priori estimation of the runtime and communication overhead of distributed SpMV (see Secs.~\ref{sec:metric},~\ref{sec:scalecomm}), (ii) the communication overhead of SpMV in the standard stack layout is large even for matrices with very regular sparsity patterns, such that consideration of alternative vector layouts is worthwhile for many large-scale computations (see Sec.~\ref{sec:panelspeedup}), (iii) the speedup by switching to a panel layout is large enough to compensate the additional communication required for global redistribution of vectors (see Sec.~\ref{sec:distribute},~\ref{sec:amortize}).
However, improved performance comes at the price of increased memory requirements for SpMV with explicit matrix storage, as discussed in Sec.~\ref{sec:memory}.

For the convenience of the reader, Table~\ref{tab:symbols} lists the most important symbols used in the following analysis.

\begin{table}
\caption{List of most important symbols used in the performance analysis in Sec.~\ref{sec:layers}.}
\label{tab:symbols}
\begin{tabular}{ll}
\toprule
Symbol & Description \\
\midrule
$D$ & matrix dimension \\
$N_s$ & number of search vectors \\
$N_t$ & number of target eigenvalues and eigenvectors\\
$\mathcal P$ & total number of processes \\
$N_\mathrm{row}$, $N_\mathrm{col}$ & number of rows and columns of Cartesian process grid \\
$N_p$ & number of processes used for one SpMV \\
$n_b$ & number of column vectors in SpMV \\
$n_\mathrm{nzr}$ & average number of non-zeros per matrix row \\
$S_d, S_i$ & size of numerical and index data type in bytes \\
$b_m$, $b_c$ & effective memory and communication bandwidth \\
$\chi_1$, $\chi_2$, $\chi_3$ & a priori communication metrics \\
$\Pi$ & parallel efficiency \\
$\kappa$ & effective factor of vector reads and writes from and to memory \\
$s$ & speedup of a single SpMV in panel layout relative to stack layout \\
$S$ & speedup of multiple SpMVs, including redistribution of vectors \\
\bottomrule
\end{tabular}
\end{table}

\subsection{Communication metrics for
distributed sparse matrix-vector multiplication}
\label{sec:metric}

In distributed SpMV, a process
 evaluates the matrix-vector product $y = A \mkern1mu x$ for several consecutive rows $i \in [a{:}b)$, as in the expression
\begin{equation}\label{SpMVprocess}
y_i = \sum_{\substack{j=0 \\A_{ij} \ne 0}}^{D-1} \mkern1mu A_{ij} \mkern1mu x_j  \qquad\qquad (\text{for } i \in [a{:}b)) \;.
\end{equation}
Here, $[a{:}b) = \{i \in \mathbb Z\,: a \le i < b\}$ denotes the range of integers from $a$ (included) to $b$ (excluded) in slice notation.
Note that we use zero-based indexing throughout this work.
If the SpMV is distributed uniformly across $N_p$ processes 
we have $b-a \approx D/N_p$.

During evaluation of Eq.~\eqref{SpMVprocess},
\begin{equation}
 n_\mathrm{m} = \big|\big\{ (i,j) : A_{ij} \ne 0 \text{ for } i \in [a{:}b), \, j \in [0{:}D) \big\}\big| 
\end{equation}
matrix elements are read from local memory. Here, $|\mathcal S|$ denotes the number of elements in the set $\mathcal S$.
Usually, $n_\mathrm{m} \approx (b-a) n_\mathrm{nzr}$, where $n_\mathrm{nzr}$ is the average number of non-zeros per matrix row.

Vector entries are distributed across the processes according to the distribution of the matrix rows.
Therefore, a lower limit for the number of vector entries read from local 
memory during evaluation of Eq.~\eqref{SpMVprocess} 
is given by
\begin{equation}
n_\mathrm{vm} = \big|\big\{ j : A_{ij} \ne 0 \text{ for } i,j \in [a{:}b) \big\}\big| \;.
\end{equation}
In many situations, especially whenever $A$ has a full diagonal, we have $n_\mathrm{vm} = b-a \approx D/N_p$.

In FD, the SpMV is executed as a block operation on $n_b$ vectors (where $n_b \approx N_s/N_\mathrm{col}$ in the panel layout).
A lower limit for the data volume read from local memory (matrix elements and vector entries) thus is
\begin{equation}\label{traffic}
V_m = n_\mathrm{m} (S_d + S_i) + (b-a) S_i + n_b n_\mathrm{vm} S_d \;,
\end{equation}
where $S_d$ (or $S_i$) denotes the size of a single matrix/vector data (or matrix index) element in bytes.
For a real (complex) matrix in double precision we have $S_d = 8$ ($S_d = 16$),
and $S_i = 4$ for \texttt{uint32} indices for matrix dimensions $D < 2^{32} \approx 4 \times 10^9$.
The second summand in Eq.~\eqref{traffic} accounts for reading of the row pointers of a matrix stored in, e.g., Compressed Sparse Row (CSR) format~\cite{Templates}, and can often be neglected.

For optimal memory access patterns and cache reuse, the matrix of vectors $V$ should be stored in row-major order~\cite{10.1007/978-3-319-92040-5_17}.
Note also that the performance of block SpMV generally improves with increasing blocksize $n_b$ since each matrix element has to be loaded only once from memory and can be reused for all block vector entries with the same row index. On the other hand, vector entries may have to be reread from main memory due to capacity misses.
These effects can be accounted for by a slight modification of Eq.~\eqref{traffic}, which is relevant for the node-level performance analysis of SpMV~\cite{Kreutzer14} but not for the communication characteristics.

The number of vector entries fetched from remote processes during evaluation of Eq.~\eqref{SpMVprocess} is 
\begin{equation}
n_\mathrm{vc} = \big|\big\{ j  : A_{ij} \ne 0  \text{ for } i \in [a{:}b), j \not\in [a{:}b) \big\}\big|  \;,
\end{equation}
which results in the inter-process communication volume
\begin{equation}
V_c = n_b n_\mathrm{vc} S_d \;.
\end{equation}
$n_\mathrm{vc}$ varies strongly with the sparsity pattern and cannot be replaced by an average value.

With memory traffic $V_m$ and communication volume $V_c$, the execution time of SpMV is 
\begin{equation}\label{tSPMVM}
 t = \max_\text{processes} \Big( \frac{V_m}{b_m} + \frac{V_c}{b_c} \Big) \;,
\end{equation}
where $b_m$  is the memory bandwidth and $b_c$ is the (effective) inter-process communication bandwidth.
Note that $b_m, b_c$ are measured per process: If multiple processes share resources, for example the main memory interface on the same node, the bandwidth must be scaled accordingly.

For perfect overlap of local memory accesses and communication, the sum in Eq.~\eqref{tSPMVM} can be replaced by $\max \{ V_m/b_m \, , V_c/b_c \}$. Perfect overlap is usually a too optimistic assumption, since also communication involves memory traffic due to \mpi{} buffer access.

Note further that in Eq.~\eqref{tSPMVM} we assume that the communication volume is large enough to ignore network latency, but not so large to result in network congestion with significantly reduced throughput.
These assumptions are in accordance with the benchmark results given below.

\paragraph{Communication metrics}

To quantify the impact of communication on distributed SpMV we introduce three essentially equivalent metrics.
The metrics are computed directly from the matrix sparsity pattern, prior to any benchmarks. 
In this way, they characterize the matrix rather than the execution time of a sparse matrix-vector operation which depends on additional factors such as the memory and communication bandwidth or the blocksize (see Eq.~\eqref{tSPMVM} for SpMV or Eq.~\eqref{EstimateT} for the Chebyshev filter).
Note that the metrics depend on how the matrix rows are distributed across the $N_p$ processes. We here give the metrics for a uniform distribution, as specified after Eq.~\eqref{SpMVprocess}, such that they depend only on $N_p$\@. All metrics assume the value zero for $N_p=1$.

The first metric is the ratio between remote and local memory accesses,
\begin{equation}\label{chi}
\chi_1 = \max_\text{processes} \quad  \frac{n_\mathrm{vc}}{n_\mathrm{vm}} \;.
\end{equation}
It distinguishes memory-bound from communication-bound SpMVs: For $\chi_1 \gg b_c/b_m$ the execution time of the SpMV according to Eq.~\eqref{tSPMVM} is dominated by communication, for $\chi_1 \ll b_c/b_m$ by local memory access.

The second metric quantifies the communication volume of SpMV relative to the vector dimension $D$,
\begin{equation}
\chi_2 = \sum_\text{processes} \quad \frac{n_\mathrm{vc}}{D} \;.
\end{equation}
We can interpret this metric in the way that after approximately $\chi_2^{-1}$ SpMVs the equivalent of a full vector has been communicated.

The third metric
\begin{equation}
\chi_3 = N_p \times \max_\text{processes} \quad \frac{n_\mathrm{vc}}{D}
\end{equation}
quantifies how communication limits the parallel efficiency of SpMV.
The parallel efficiency $\Pi = t[1]/(N_p \mkern2mu t[N_p])$ measures how the execution time $t[N_p]$ scales with the number of processes $N_p$.
For ideal scaling, where $t[N_p] = t[1]/N_p$, we have $\Pi = 1$, otherwise $\Pi < 1$.

In terms of the metric $\chi_3$, we can formulate the estimate 
\begin{equation}\label{PiParallel}
 \Pi \lesssim \min\{1, \chi_3^{-1} \frac{b_c}{b_m} \} \;.
\end{equation}
To see why, note that according to Eqs.~\eqref{traffic},~\eqref{tSPMVM} the execution time on a single compute process is $t[1] = V_m/b_m = (n_b S_d / b_m) \times D$ (if we neglect data traffic from matrix elements for large $n_b$ and assume $n_\mathrm{vm} = D$).
For distributed SpMV on $N_p>1$ processes we get the ideal speedup $t[1]/t[N_p] = N_p$ if no communication is required (and if $n_\mathrm{vm}$ scales as $D/N_p$, which is a safe assumption).
With communication the execution time is at least $t[N_p] \ge V_c/b_c = (n_b S_d /b_c) \times n_\mathrm{vc}$, which results in the reduced parallel efficiency $\Pi = t[1]/(N_p \mkern2mu t[N_p]) \le (b_c/b_m) \times D/(N_p n_\mathrm{vc})$. With the definition of $\chi_3$ we arrive at the given bound on $\Pi$.

Note that by definition the communication metrics $\chi_2$, $\chi_3$ directly give the communication volume per process.
The metric $\chi_2$ gives the average communication volume per process
\begin{equation}\label{VCAve}
 V_c^\mathrm{ave} = \frac{n_b D S_d}{N_p} \, \chi_2 \;,
\end{equation}
the metric $\chi_3$ the maximal communication volume per process
\begin{equation}\label{VCMax}
 V_c^\mathrm{max} = \frac{n_b D S_d}{N_p} \, \chi_3 \;,
\end{equation}
both for a SpMV with $n_b$ vectors of length $D$ and a data type of size $S_d$. Note further that these values are known already prior to execution of the SpMV.

Under reasonable assumptions
on the regularity of the sparsity pattern,
the three metrics $\chi_{1,2,3}$ are equivalent up to small factors
(two counterexamples are given in App.~\ref{app:irregular}).
First, recall that we almost always have $n_\mathrm{vm} \approx D/N_p$ in Eq.~\eqref{chi},
that is $\chi_1 \approx \chi_3$.
Second, unless the communication volume per process varies strongly, it does not make a difference whether we take the maximum or average over all processes. Therefore, we also have $\chi_2 = N_p \times (1/N_p) \sum_\text{processes} (n_\mathrm{vc}/D) \approx N_p \max_\text{processes} (n_\mathrm{vc}/D) = \chi_3$, but a small factor may appear in this relation. A large discrepancy between $\chi_{1,3}$ and $\chi_2$,
and thus between the
average and maximal communication volume $V_c^\mathrm{ave}$ and $V_c^\mathrm{max}$ per process according to Eqs.~\eqref{VCAve},~\eqref{VCMax},
would indicate a severe communication imbalance among the processes.
Then, a careful implementation may try to adjust the distribution of rows onto processes to equalize the communication volume.
Note that advanced methods for communication optimization do exist~\cite{ACER201671}.

The communication metrics described above could be applied to other sparse solvers. However, most solvers
require additional components whose communication overhead has 
to be modeled separately.
For example, orthogonalization requires global reduction (hence the consideration of redistribution of vectors in Sec.~\ref{sec:distribute} below), while the Chebyshev filter requires only SpMV.

\subsection{Communication metrics of real application matrices}
\label{sec:scalecomm}

\begin{table}
\caption{Communication metrics for the \texttt{Exciton} and \texttt{Hubbard} matrices.}
\label{tab:commcomp}
\medskip
\hfill
\begin{tabular}{lrrr}
\toprule
matrix \hspace{2cm}  & $N_p$  & $\chi_{1,3}$ & $\chi_2$ \\ 
\midrule
\texttt{Exciton,L=75}  & 2 & 0.01 & 0.01 \\
 &  4 & 0.05 & 0.04 \\
 &  8 & 0.11 & 0.09 \\
\quad $D = 10\,328\,853$ &  16 & 0.21 & 0.20 \\
\quad $n_\mathrm{nzr} = 8.96$    & 32 & 0.42 & 0.41 \\
                                 & 64 & 0.85 & 0.83 \\[4ex]
\texttt{Exciton,L=200}    & 2 & 0.00 & 0.00 \\
 &  4 & 0.02 & 0.01 \\
 &  8 & 0.04 & 0.03 \\
\quad $D = 193\,443\,603$ &  16 & 0.08 & 0.07 \\
\quad $n_\mathrm{nzr} = 8.99$    & 32 & 0.16 & 0.15 \\
                                 & 64 & 0.32 & 0.31 \\
\bottomrule
\end{tabular}
\hfill{}
\begin{tabular}{lrrrrr}
\toprule
matrix  \hspace{2cm} & $N_p$  & $\chi_{1,3}$ & $\chi_2$ \\ 
\midrule
\texttt{Hubbard,n\_sites=14,}   & 2 & 0.54 & 0.54 \\
\texttt{\quad n\_fermions=7} &  4 & 1.51 & 1.02 \\
 &  8 & 2.52 & 1.53 \\
\quad $D = 11\,778\,624$ &  16 & 3.37 & 2.07 \\
\quad $n_\mathrm{nzr} = 14.00$    & 32 & 4.17 & 2.65 \\
                                  & 64 & 5.58 & 3.19 \\[4ex]
\texttt{Hubbard,n\_sites=16,}   & 2 & 0.53 & 0.53 \\
 \texttt{\quad n\_fermions=8}   &  4 & 1.50 & 1.01 \\
 &  8 & 2.50 & 1.51 \\
\quad $D = 165\,636\,900$  &  16 & 3.37 & 2.03 \\
\quad $n_\mathrm{nzr} = 16.00$   & 32 & 4.21 & 2.61 \\
                                 & 64 & 5.67 & 3.16 \\
\bottomrule
\end{tabular}
\hfill{}
\end{table}

Table~\ref{tab:commcomp} lists the communication metrics for the \texttt{Exciton} and \texttt{Hubbard} matrices from Fig.~\ref{fig:sparsity}.
The matrices are taken from the Scalable Matrix Collection \scamac{} and can be generated in various sizes controlled by the problem parameters~\cite{SCAMAC}. In \scamac{} notation, the matrices in Fig.~\ref{fig:sparsity} are called ``\texttt{Exciton,L=200}'' and ``\texttt{Hubbard,n\_sites=16,n\_fermions=8}.''

The \texttt{Exciton} matrix is essentially a 3D stencil with additional local blocks.
The communication metrics are small, which fits the ``narrow'' sparsity pattern in Fig.~\ref{fig:sparsity}.
On the other hand, the ``wide'' sparsity pattern of the \texttt{Hubbard} matrix results in large values of the communication metrics.
From the values in Table~\ref{tab:commcomp} we see that $\chi>0.5$ already for $N_p \ge 2$. Using our previous estimates we expect that communication dominates SpMV in this situation.

\begin{figure}
\includegraphics[scale=0.8]{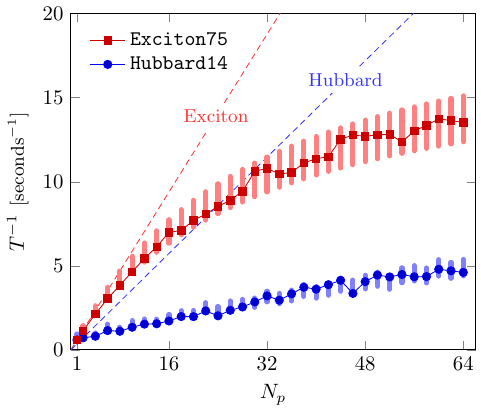}
\hspace*{\fill}
\includegraphics[scale=0.8]{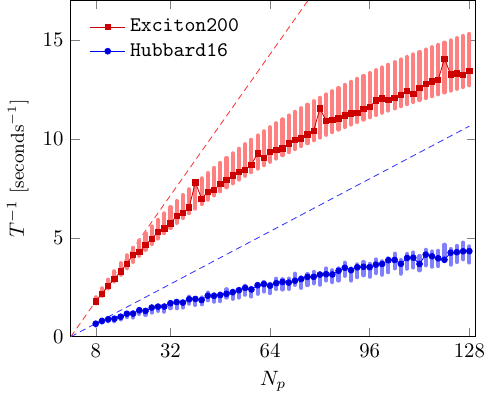}
\caption{Inverse execution time of one iteration of the Chebyshev filter (Alg.~\ref{alg:chebfilter}) as a function of the number of parallel processes. The vertical bars give the theoretical estimate according to Eq.~\eqref{EstimateT}, with a $\pm 10 \%$ variation of the model parameters $b_c, \kappa$.
The dashed line corresponds to perfect strong scaling with parallel efficiency $\Pi=1$; in the right panel, the execution time for $N_p=8$ processes is used as a baseline. The left (right) panel uses $n_b=64$ ($n_b=8$) vectors.
}
\label{fig:bench}
\end{figure}

\paragraph{Benchmarks}
To test the predictive power of the communication metrics, Fig.~\ref{fig:bench} shows benchmark results for the Chebyshev filter in Algorithm~\ref{alg:chebfilter}, with the four matrices from Table~\ref{tab:commcomp}.
The benchmarks have been executed on the ``Meggie'' cluster at Erlangen National High Performance Computing Center (NHR@FAU)\footnote{Hardware documentation is available at \\ \url{https://hpc.fau.de/systems-services/systems-documentation-instructions/clusters/meggie-cluster/}}.
Clearly, the execution time of SpMV depends on the mapping of processes onto compute nodes.
For all benchmarks, we use a NUMA-friendly ``one process per socket'' mapping~\cite{HagerWelleinBuch}.
On the Meggie cluster, this corresponds to two \mpi{} processes per node, with $10$ \openmp{} threads each.
Data in the left panel of Fig.~\ref{fig:bench} are obtained on up to $32$ nodes, and include the data point $N_p=1$ for a single process on one socket (no communication). Data in the right panel are obtained on up to $64$ nodes (i.e., $N_p\leq 128$); at least $4$ nodes are required due to memory demands for the larger matrices.

To get a quantitative idea of the communication involved in the SpMV we can either use Eqs.~\eqref{VCAve},~\eqref{VCMax} together with the metrics from Table~\ref{tab:commcomp}, or compute the communication volume directly in bytes from the sparsity pattern. Note that the \texttt{Exciton} (\texttt{Hubbard}) matrices are complex (real) valued, with sizes $S_d=16$ ($S_d=8$) of the double precision datatype. Selected values for the communication volume are given in Table~\ref{tab:commvol}.
The communication volume is certainly large enough to justify that our model neglects the latency of the \mpi{} communication, which is clearly volume bound.
Note also that the average and maximal communication volume are comparable, reflected by $\chi_2 \approx \chi_3$, unless the communication volume becomes so large that the performance of SpMV degrades in any case (compare with Fig.~\ref{fig:bench}).

\begin{table}
\caption{Average and maximal communication volume per process and SpMV for the four matrices in Fig.~\ref{fig:bench}.}\label{tab:commvol}
\medskip
\hspace*{\fill}
\begin{tabular}{lr@{\hspace{5mm}}rr}
\toprule
 & & \multicolumn{2}{c}{communication volume} \\
\smash{\raisebox{1.5ex}{matrix}} \hspace{1cm}  & \smash{\raisebox{1.5ex}{$N_p$}}  & \multicolumn{1}{r}{average} & \multicolumn{1}{r}{maximal} \\ 
\midrule
\texttt{Exciton75}
 & 2 & \SI{66.8}{\mebi\byte} & \SI{66.8}{\mebi\byte} \\
 & 64 & \SI{131.51}{\mebi\byte} & \SI{133.6}{\mebi\byte} \\[0.5ex]
\texttt{Hubbard14}
 & 2 & \SI{1.51}{\gibi\byte} & \SI{1.51}{\gibi\byte} \\
 & 64 & \SI{286.33}{\mebi\byte} & \SI{501.27}{\mebi\byte} \\[0.5ex]
\texttt{Exciton200}
 & 4 & \SI{88.33}{\mebi\byte} & \SI{117.73}{\mebi\byte} \\
 & 64 & \SI{115.93}{\mebi\byte} & \SI{117.77}{\mebi\byte} \\[0.5ex]
\texttt{Hubbard16}
 & 4 & \SI{2.5}{\gibi\byte} & \SI{3.69}{\gibi\byte} \\
 & 64 & \SI{499.79}{\mebi\byte} & \SI{895.86}{\mebi\byte} \\
\bottomrule
\end{tabular}
\hspace*{\fill}
\end{table}

We can now compare the benchmark results to the theoretical prediction of the model in Eq.~\eqref{tSPMVM} for
the execution time of one iteration of the Chebyshev filter on $N_p$ processes and $n_b$ vectors,
\begin{equation}\label{EstimateT}
 T = \Bigg[ \Big(\frac{(S_d+S_i) n_\mathrm{nzr}}{n_b} + \kappa S_d \Big) \frac{1}{b_m} \, + \, \chi[N_p]  \frac{S_d}{b_c} \Bigg] \times n_b \times \frac{D}{N_p} \;.
\end{equation}
This expression has three model parameters $b_m, b_c, \kappa$,
while $\chi[N_p]$ is not a parameter but computed from the matrix sparsity pattern as in Table~\ref{tab:commcomp}.
Generally, the blocksize $n_b$ is large enough (we always have $n_b \ge 8$) to allow us to neglect the first term for the memory traffic from the matrix at least in the qualitative estimates given later.

To determine the model parameters we can measure the bare memory bandwidth $b_m$ with the STREAM benchmark~\cite{McCalpin95}.
The parameter $b_c$, on the other hand, is an effective communication bandwidth that has to account for all communication-related operations in the SpMV, including the copying of halo elements to \mpi{} buffers and other \mpi{} overhead. 
On typical compute clusters we can expect a factor $b_m/b_c^\text{bare} \approx 10 \dots 20$ between the memory bandwidth and the bare communication bandwidth $b_c^\text{bare}$ of the cluster interconnect.
Apart from highly idealized situations, the value of the effective $b_c$ cannot be derived from $b_c^\text{bare}$. Instead, we use a single value of $b_c$ for each curve in Fig.~\ref{fig:bench}, which is obtained from a fit to the benchmark data. Including the communication-related overhead in SpMV a factor $b_c^\text{bare}/b_c$ in the range $1 \dots 2$ appears to be a reasonable fit result.

The factor $\kappa$ in Eq.~\eqref{EstimateT}  accounts for reading and writing of the vectors $W_1, W_2, V$ from and to memory in Algorithm~\ref{alg:chebfilter}.
With ideal cache usage, we have $\kappa=5$ since $W_1$ is read once from memory, and $W_2, V$ are read and written.
If instead of a fused kernel the two operations in step 7 are executed independently, the factor increases to $\kappa=6$ since $W_2$ has to be read twice (for large problems it is likely that $W_2$ has been evicted from cache after the SpMV).
For an isolated SpMV, without the \texttt{axpby} operation, the factor would reduce to $\kappa = 3$.

In practice, cache usage is never ideal but vector entries have to be read more than once from memory due to capacity misses~\cite{Kreutzer14},
resulting in $\kappa > 5$. The precise value depends on a number of conditions, especially how regular or scattered the memory access is.
We determine the parameter $\kappa$ from the benchmark for a single process ($N_p=1$) and use this value for all $N_p \ge 1$. Although this is an oversimplification since $\kappa$ changes with the distribution of the matrix onto the processes, this choice agrees sufficiently well with the benchmark data.

The values of the model parameters for the four matrices in Fig.~\ref{fig:bench} are given in Table~\ref{tab:model}.
The values are consistent with our basic assumptions.
First, we have $\kappa > 5$, and $\kappa$ is larger for the Hubbard matrix with the more irregular memory access.
Second, the effective communication bandwidth $b_c$ lies consistently within a $\pm 10\%$ range across the four matrices,
and the ratio $b_m/b_c \approx 15 \dots 20$ is compatible with our expectations for the effective communication bandwidth of our SpMV. 

Note that once $b_m, b_c, \kappa$ are fixed in Eq.~\eqref{EstimateT}, the entire dependence of $T$ on $N_p$ arises from the communication metric $\chi[N_p]$, which is computed independently of the model parameters or benchmark data.
As Fig.~\ref{fig:bench} shows, the theoretical prediction agrees very well with the benchmark data for all values of $N_p$,
which validates our model.

Further evidence for our model is that all terms in Eq.~\eqref{EstimateT} are required for agreement between model and benchmarks. 
Without the first term from local memory traffic the model would predict $T=0$ for $N_p=1$ since $\chi[1]=0$ (no communication for a single process).
This is of course not correct.
Without the explicit dependence of the communication metric $\chi$ on $N_p$ in the second term the model would predict perfect scaling $T \propto 1/N_p$. This is not correct as shown by the dashed lines in Fig.~\ref{fig:bench}. In particular we cannot replace $\chi[N_p]$ with a constant.

\begin{table}
\caption{Model parameters $b_m$, $\kappa$, $b_c$ for the data in Fig.~\ref{fig:bench}. Values marked with a $*$ are taken from the row(s) immediately above.  Especially the value of $b_m$ (from the STREAM benchmark) is the same for all matrices.}
\label{tab:model}
\medskip
\hspace*{\fill}
\begin{tabular}{lccc}
\toprule
\multicolumn{1}{c}{matrix} &\rule{1ex}{0pt} $b_m$ [\si{\giga\byte\per\second}] \rule{1ex}{0pt} & $\kappa$ & $b_c$ [\si{\giga\byte\per\second}] \\
\midrule
\texttt{Exciton75} &  53.3 & 7.30   & 2.82 \\
\texttt{Exciton200} & $*$ & $*$ & 3.10 \\[0.5ex]
\texttt{Hubbard14} & $*$ & 10.0 & 2.82 \\
\texttt{Hubbard16} & $*$ & $*$ & 2.54 \\
\bottomrule
\end{tabular}
\hspace*{\fill}
\end{table}
The behavior of the communication metrics in Table~\ref{tab:commcomp} and of the benchmark data in Fig.~\ref{fig:bench} is not specific to the \texttt{Exciton} or \texttt{Hubbard} matrices but of general nature.
Appendix~\ref{app:scamac} lists communication metrics and benchmark results for several other matrices from \scamac{}.
In all cases, our theoretical model agrees very well with the benchmark data.
We also include results for matrices from the SuiteSparse (former: Florida) matrix collection~\cite{SuiteSparse,Florida}, which are rather different from the typical (quantum) physics matrices of main interest for the present study, but still comply with our model.
In short, the appendix shows that matrices with significant communication overhead in SpMV, reflected by large values of the communication metrics, are not uncommon.

\subsection{Panel layout speedup}
\label{sec:panelspeedup}

So far, we have analyzed the execution time  of the Chebyshev filter as a function of the number $N_p$ of processes used for the parallel evaluation of the SpMV. In terms of vector layouts (see Fig.~\ref{fig:layouts}) the analysis applies to the standard stack layout where the sparse matrix rows, hence the vectors, are distributed equally across all the $N_p$ processes.

We now turn to the concept of two orthogonal layers of parallelism.
Our goal is to use Eq.~\eqref{EstimateT} to estimate the speedup that can be achieved if the Chebyshev filter is evaluated in a panel layout instead of the stack layout.
Recall that we can vary the shape of the panel layout provided that the total number $\mathcal P = N_\mathrm{row} \times N_\mathrm{col}$ of processes remains fixed.

In a panel layout the SpMVs in the Chebyshev filter are executed simultaneously and independently in each of the process columns. Each SpMV uses a fraction $N_p = \mathcal P/N_\mathrm{col} = N_\mathrm{row}$ of all processes and is applied to $n_b = N_s / N_\mathrm{col}$ vectors. 
The stack layout corresponds to $N_\mathrm{col}=1$, with $N_p = \mathcal P$ and $n_b = N_s$.
The two parameters in Eq.~\eqref{EstimateT} that change between the stack and panel layout thus are $N_p$ and $n_b$. With these parameters also the value of the communication metric  $\chi[N_p]$ changes. Note that the last factor $n_b \times (D/N_p) = N_s D / \mathcal P$ in Eq.~\eqref{EstimateT} is constant.

With the respective parameter values
the execution time of the Chebyshev filter is 
\begin{equation} 
 T_\mathrm{stack} = T[\mathcal P, N_s] = S_d \times \bigg( \frac{\kappa}{b_m}  +   \frac{\chi[\mathcal P]}{b_c} \bigg) \times \frac{N_s D}{\mathcal P}
\end{equation} 
  in the stack layout and 
\begin{equation}\label{Tpanel}
  T_\mathrm{panel} = T\bigg[\frac{\mathcal P}{N_\mathrm{col}}, \frac{N_s}{N_\mathrm{col}}\bigg] =  S_d \times \bigg( \frac{\kappa}{b_m}  +  \frac{\chi[\mathcal P/N_\mathrm{col}]}{b_c} \bigg) \times \frac{N_s D}{\mathcal P}
\end{equation}   
in the panel layout.
For the sake of clarity, we omit in both expressions the memory traffic from the matrix elements (this is the first term in Eq.~\eqref{EstimateT}), which is justified for reasonable values of $n_b$. 

The speedup of the panel relative to the stack layout is
\begin{equation}\label{speedup}
s =  \frac{T_\mathrm{stack}}{T_\mathrm{panel}} = \frac{\kappa \mkern1mu (b_c/b_m) +   \mkern1mu \chi[\mathcal P]}{\kappa  \mkern1mu (b_c/b_m) + \chi[\mathcal P/N_\mathrm{col}]} \;.
\end{equation}
In the communication-bound regime $\chi \gg \kappa (b_c/b_m)$, the speedup is given by
\begin{equation}\label{scommbound}
s \simeq \frac{\chi[\mathcal P]}{\chi[\mathcal P/N_\mathrm{col}]} 
\end{equation}
and depends entirely on the communication metric.

\begin{figure}
\hspace*{\fill}
\includegraphics[scale=0.8]{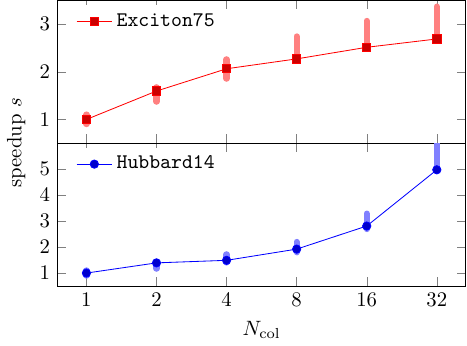}
\hspace*{\fill}
\includegraphics[scale=0.8]{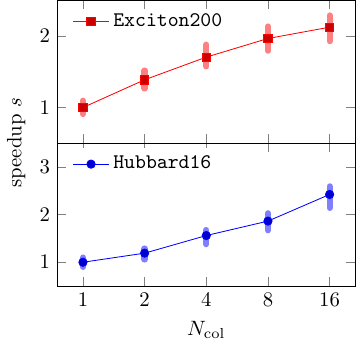}
\hspace*{\fill}
\caption{Speedup of the Chebyshev filter in the panel layout relative to the stack layout, as a function of the number of process columns $N_\mathrm{col}$
($N_\mathrm{col} = 1$ gives the stack layout, hence $s=1$). Data in the left (right) panel are obtained 
on $32$ ($64$) nodes. 
For the larger matrices in the right panel a minimum of four nodes is required due to memory demands from storing the matrix,
hence $N_\mathrm{col} \le 16$. 
}
\label{fig:bench2}
\end{figure}

Figure~\ref{fig:bench2} shows benchmark results for the speedup $s$ for different shapes of the panel layout, in comparison to the theoretical prediction from Eq.~\eqref{speedup}. Again, we find very good agreement. Notably, even for the exciton matrices with the ``narrow'' sparsity pattern and small values $\chi < 1$ of the communication metric a significant speedup $s > 2$ can be observed.

Note that the maximal speedup 
is achieved for $N_\mathrm{col}=\mathcal P$, i.e., for the pillar layout,
where SpMV does not require communication. If considered in isolation, the pillar layout would result in perfect (weak and strong) scaling of the Chebyshev filter with parallel efficiency $\Pi[\mathrm{pillar}] = 1$.
However, in FD the Chebyshev filter is used in conjunction with orthogonalization, which cannot be performed in a scalable or efficient way in the pillar layout since entire vectors would have to be communicated.
Furthermore, the memory overhead of the pillar layout must be taken into account if a matrix-free
formulation is not available and the matrix has to be stored redundantly.

\subsection{Redistribution of vectors between the panel and stack layout}
\label{sec:distribute}

Orthogonalization of vectors in the panel layout requires communication of vector entries along the horizontal axis of Fig.~\ref{fig:layouts}.
Since every vector has to be matched against every other vector, the aggregated communication volume is at least of order $N_s D (1- 1/N_\mathrm{col})$ vector entries.
It is not obvious how to organize the ``all-to-all'' communication of vectors in the panel layout,
especially since memory consumption becomes problematic if vectors are gathered at one (or at every) process.

Because of these difficulties we do not orthogonalize vectors in the panel layout but instead redistribute vectors between the panel and stack layout.
Redistribution of vectors is a global operation with a controlled communication pattern.
The communication volume is of the same order as for (the hypothetical) orthogonalization in the panel layout.
Redistribution can be performed in-place, and does not require additional memory.
Once vectors are in the stack layout, the known communication-avoiding orthogonalization techniques can be used.

\begin{figure}
\hfill{}
\raisebox{0\height}{\includegraphics[scale=0.9]{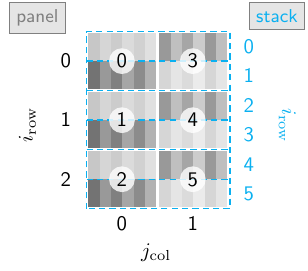}}
\hfill{}
\raisebox{0.5\height}{\includegraphics[scale=0.9]{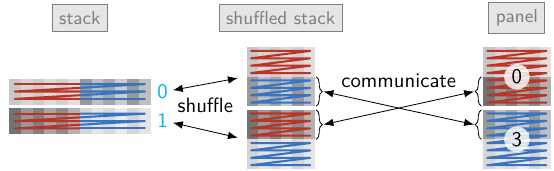}}
\hfill{}
\caption{Left: $3 \times 2$ panel vs stack layout, with process ranks assigned in column-major order to the Cartesian grid of the panel.
Right:
Redistribution of vectors between the stack and panel layout requires shuffling of the stack (a local operation without communication) to preserve contiguous memory storage of vector entries (which are in row-major order, indicated by the zigzag lines) and horizontal communication of vector entries between processes. 
For the matching $3 \times 2$ layout shown here, communication occurs only between the two processes in the same row of the panel.
}
\label{fig:RedistributePattern}
\end{figure}

Consider $N_s$ vectors $V_0, \dots, V_{N_s-1}$ with dimension $D$ that are distributed across $\mathcal P$ processes once in the stack, once in the panel layout (see Fig.~\ref{fig:RedistributePattern}).
As before, the vectors are arranged as the columns of a $D \times N_s$ matrix $V$,
such that $V_{ij}$ is the $i$-th entry of the $j$-th vector $V_j$.

The stack layout is specified by row indices $k_0, \dots, k_{\mathcal P}$ that slice the range $[0{:}D)$,
such that $0 = k_0 \le k_1 \le \dots \le k_{\mathcal P-1} \le k_{\mathcal P} = D$.
Process $i_\mathrm{row}$, with index $i_\mathrm{row} \in [0{:}\mathcal P)$, stores a horizontal slice of the matrix $V$, given by the vector entries $V_{ij}$ with $i \in [k_{i_\mathrm{row}}{:}k_{i_\mathrm{row}+1})$ and $j \in [0{:}N_s)$. These are $N_s (k_{i_\mathrm{row}+1} - k_{i_\mathrm{row}})$ vector entries in total, that is about $N_s D/\mathcal P$ for nearly equidistant row indices.

The $N_\mathrm{row} \times N_\mathrm{col}$ panel layout 
is specified by row indices $m_0, \dots, m_{N_\mathrm{row}}$ that slice the range $[0{:}D)$ and column indices $n_0, \dots, n_{N_\mathrm{col}} $ that slice the range $[0{:}N_s)$.
Process $(i_\mathrm{row}, j_\mathrm{col})$ 
stores a rectangular part of the matrix $V$, given by the vector entries $V_{ij}$ with $i \in [m_{i_\mathrm{row}}{:}m_{i_\mathrm{row}+1})$ and $j \in [n_{j_\mathrm{col}}{:}n_{j_\mathrm{col}+1})$.
These are $(m_{i_\mathrm{row}+1} - m_{i_\mathrm{row}}) (n_{i_\mathrm{col}+1} - n_{i_\mathrm{col}})$ vector entries in total, that is about $(D/N_\mathrm{row})(N_s/N_\mathrm{col}) = N_s D/ \mathcal P$ for nearly equidistant row and column indices. This number agrees with the number of vector entries in the stack layout.

To redistribute vectors between the stack and panel layout we have to communicate vector entries horizontally in the graphical representation of Fig.~\ref{fig:RedistributePattern}.
Redistribution of vectors from the stack to the panel layout and from the panel to the stack layout are inverse operations.
The communication patterns are identical, only the direction of communication changes.

Most communication takes place between processes in the same row of the panel layout.
For matching layouts, where exactly $m_{i_\mathrm{row}} = k_{(i_\mathrm{row} \times N_\mathrm{col})}$, communication strictly takes place between processes in the same row only.
The general communication pattern for arbitrary choices of row and column indices involves technical details of minor relevance and is omitted here.
In our implementation, communication is performed with a collective \texttt{MPI\_Alltoall} operation.
Redistribution also requires shuffling of vector entries to preserve contiguous memory storage during and after the \mpi{} operations.
For row-major order as depicted in Fig.~\ref{fig:RedistributePattern}, we shuffle vector entries in the stack layout, for column-major order we would shuffle vector entries in the panel layout.
Shuffling is a strictly local operation without communication.

Note that in Fig.~\ref{fig:RedistributePattern} we assign process ranks in column-major order to the Cartesian grid of the panel layout.
This choice puts ``adjacent processes'' with nearby rank into the same column of the panel,
to acknowledge the fact that SpMV involves communication along the process columns.
In any case, the process indices $(i_\mathrm{row}, j_\mathrm{col})$ in the panel layout, or $i_\mathrm{row}$ in the stack layout, are logical indices rather than physical ranks and can be assigned freely.
Which assignment is most efficient depends on the network topology.

To quantify the communication volume of redistribution we note that a process has to exchange all vector entries apart from those that lie in overlapping regions of the two layouts.
In Fig.~\ref{fig:RedistributePattern}, the top left rectangle and the bottom right rectangle in each process row (colored in light gray in the figure) overlap and do not need to be communicated. The height (or width) of such a region is approximately $D/\mathcal P$ (or $N_s/N_\mathrm{col}$).
Therefore, the communication volume per process is approximately
$N_s D/\mathcal{P} - (D/\mathcal P) (N_s/N_\mathrm{col}) =  (N_s D/\mathcal{P}) (1-1/N_\mathrm{col})$ vector entries.
Summation over all columns gives the communication volume
\begin{equation}\label{volredist}
\frac{V_\mathrm{row}}{S_d} = N_s D \frac{N_\mathrm{col}-1}{\mathcal P}  = N_s \frac{D}{N_\mathrm{row}}  \, (1- 1/N_\mathrm{col})
\end{equation}
per process row, and
the total communication volume
\begin{equation}\label{volredistfull}
\frac{V}{S_d} = N_s D (1- 1/N_\mathrm{col}) \;.
\end{equation}
These values are exact for matching layouts, and reproduce our initial estimate.

\subsection{Amortization of redistribution of vectors to the panel layout}
\label{sec:amortize}

For the Chebyshev filter, 
Fig.~\ref{fig:bench} shows how scaling is lost in the stack layout due to communication,
while Fig.~\ref{fig:bench2} shows how parallel efficiency can be regained in the panel layout.
In FD, the Chebyshev filter does not occur in isolation but in conjunction with orthogonalization. Scaling and parallel efficiency of FD thus depend both on the speedup of the Chebyshev filter in the panel layout and the additional time for redistribution of vectors between the stack and panel layout.

To evaluate the Chebyshev filter in the panel layout and perform orthogonalization in the stack layout we have to redistribute 
vectors twice (steps 7,\,9 in Alg.~\ref{alg:peigfex}).
Whether the cost of redistribution is amortized by the speedup of the Chebyshev filter depends on three numbers:
(i) the speedup $s = T_\mathrm{stack} / T_\mathrm{panel} > 1$ of a single iteration of the Chebyshev filter in the panel layout,
(ii) the execution time $T_\mathrm{rd}$ of redistribution, which we can express in terms of Chebyshev filter iterations by the factor $r = T_\mathrm{rd}/T_\mathrm{panel} > 1$ (that is, one redistribution corresponds to $r$ SpMVs in the panel layout),
(iii) the number of iterations in the Chebyshev filter, given by the degree $n$ of the filter polynomial.

Combining the three numbers, the speedup of the Chebyshev filter in the panel layout relative to the stack layout, now including the time for redistribution, is
\begin{equation}\label{BigS}
S = \frac{n T_\mathrm{stack}}{n T_\mathrm{panel} + 2 T_\mathrm{rd}} = s \frac{n}{n + 2 r} \;.
\end{equation}
Therefore, switching to the panel layout becomes favorable (i.e., $S > 1$) for 
\begin{equation}\label{Favorable}
 n > n^* = \frac{2 r}{s-1} \;.
\end{equation}
The asymptotic speedup for $n \gg n^*$ is $S \simeq s$.

We can now make the connection to the communication metric $\chi$ (with $\chi \approx \chi_2 \approx \chi_{1,3}$).
The speedup $s$ is already given in Eq.~\eqref{speedup}.
For the factor $r$ we compare the communication volume of redistribution in Eqs.~\eqref{volredist},~\eqref{volredistfull} with the execution time  according to Eq.~\eqref{Tpanel} and find
\begin{equation}
 r = \frac{1-1/N_\mathrm{col}}{\kappa \, b_c/b_m + \chi[\mathcal P/N_\mathrm{col}]}  \;.
 \end{equation}
These values can be used in Eqs.~\eqref{BigS},~\eqref{Favorable}.
In the communication-bound regime $\chi \gg \kappa (b_c/b_m)$, we have 
$s \simeq \chi[\mathcal P] / \chi[\mathcal P/N_\mathrm{col}]$ as in Eq.~\eqref{scommbound} and
$r \simeq (1-1/N_\mathrm{col})/\chi[\mathcal P/N_\mathrm{col}]$
(recall our interpretation of $\chi$ that the equivalent of a full vector is communicated every $\chi^{-1}$ SpMVs).
Our analysis thus predicts, up to small factors, that it is favorable to switch to a panel layout with $N_\mathrm{col} > 1$ if
\begin{equation}
 n > \frac{2(1-1/N_\mathrm{col})}{\chi[\mathcal P] - \chi[\mathcal P/N_\mathrm{col}]} \;.
\end{equation}
For larger values of $\chi[\mathcal P]$ the condition is fulfilled already for smaller values of $n$.

For the pillar layout, with $N_\mathrm{col} = \mathcal P$, the condition simplifies to
\begin{equation}
 n_\text{[pillar]} \ge \frac{2}{\chi[\mathcal P]} \;,
\end{equation}
since $\chi[1]=0$ and  $(1-1/N_\mathrm{col}) < 1$.
This condition is fulfilled for \emph{any} $n \ge 1$ if $\chi[\mathcal P] \ge 2$.
For the \texttt{Hubbard} matrices this is the case already for $\mathcal P \ge 16$ (see Table~\ref{tab:commcomp}).
In this situation it is \emph{always} favorable to switch to a pillar layout for evaluation of the Chebyshev filter, even for $n=1$.

\begin{table}
\caption{Amortization of redistribution of vectors. The numbers are for $\mathcal P = 32$ ($\mathcal P=64$) processes for the \texttt{Exciton75} and \texttt{Hubbard14} (\texttt{Exciton200} and \texttt{Hubbard16}) matrices. For the larger matrices, the available memory required to store the sparse matrix restricts the data to $N_\mathrm{row} \ge 4$, that is $N_\mathrm{col} \le 16$.
The additional data for $N_\mathrm{col} = 64^*$ are obtained with the matrix-free SpMV used for the FD computations in Sec.~\ref{sec:FDlayers}, which has slightly lower node-level performance.
}
\label{tab:amortize}
\medskip
\hspace*{\fill}
\scalebox{0.95}{
\begin{tabular}{cc@{\hspace{4ex}}ccc@{\hspace{6ex}}ccccc}
\toprule
 & &  & & & \multicolumn{5}{c}{speedup $S$ for $n$ SpMVs} \\[0.25ex]
matrix & $N_\mathrm{col}$ & $s$ & $r$ & $n^*$ & $n=10$ & $n=20$ & $n=30$ & $n=50$ & $n=100$ \\
\midrule
\texttt{Exciton75} & 2 & $1.60$ & 4 & 14 & --- & $1.14$ & $1.26$ & $1.38$ & $1.48$
\\
\rule{5ex}{0.5pt} & 8 & $2.27$ & 8 & 13 & --- & $1.26$ & $1.48$ & $1.72$ & $1.96$
\\
\rule{5ex}{0.5pt} & 32 [pillar] & $2.69$ & 9 & 11 &  --- & $1.42$ & $1.68$ & $1.98$ & $2.28$ \\[1ex]
\texttt{Hubbard14} & 2 & $1.39$ & 1 & 6 &
$1.16$ & $1.26$ & $1.30$ & $1.34$ & $1.36$
\\
\rule{5ex}{0.5pt} & 8 & $1.92$ & 2 & 5 &
$1.37$ & $1.60$ & $1.69$ & $1.78$ & $1.85$
\\
\rule{5ex}{0.5pt} & 32 [pillar] & $4.98$ & 4 & 2 &
$2.77$ & $3.56$ & $3.93$ & $4.29$ & $4.61$
\\[1ex]
\texttt{Exciton200} & 2 & $1.39$ & 17 & 87 & --- & --- & --- & --- & $1.04$ \\
\rule{5ex}{0.5pt} & 8 & $1.97$   & 27 & 56 &  --- & --- & --- & --- & $1.28$\\
\rule{5ex}{0.5pt} & 16 & $2.13$  & 31 & 54 & --- & --- & --- & --- & $1.31$  \\
\rule{5ex}{0.5pt} & $64^*$ [pillar$^*$] & $2.02$  & 29 & 56 & --- & --- & --- & --- & $1.28$  \\[1ex]
\texttt{Hubbard16} & 2 & $1.19$  & 2 & 21 & --- & --- & $1.05$ & $1.10$ & $1.14$ \\
\rule{5ex}{0.5pt} & 8 & $1.86$   & 4 & 9 & $1.03$ & $1.33$ & $1.47$ & $1.60$ & $1.72$ \\
\rule{5ex}{0.5pt} & 16 & $2.42$  & 5 & 7 &  $1.21$ & $1.61$ & $1.81$ & $2.02$ & $2.20$ \\
\rule{5ex}{0.5pt} & $64^*$ [pillar$^*$] & $7.25$  & 15 &  5 & $1.81$ & $2.90$ & $3.62$ & $4.53$ & $5.58$  \\
\bottomrule
\end{tabular}
}
\hspace*{\fill}
\end{table}

Numerical values for our example matrices are collected in Table~\ref{tab:amortize}.
The two empirical values in the table measured in the benchmarks are the speedup of the Chebyshev filter, identical to Fig.~\ref{fig:bench2}, and the factor $r$ quantifying redistribution of vectors.
From these two values we compute the break even point $n^*$ according to Eq.~\eqref{Favorable} and the speedup $S$ for $n > n^*$ according to Eq.~\eqref{BigS}.
Since we now include the time for redistribution of vectors, the speedup $S$ is smaller than the speedup $s$ of the isolated Chebyshev filter. For a large number of SpMVs ($n \gg n^*$), the speedup reaches its maximal value ($S \to s$). Especially in the pillar layout $N_\mathrm{col}=\mathcal P$ this coincides with the parallel efficiency reaching the optimum value $\Pi \to 1$.

Recall that for the isolated Chebyshev filter the pillar layout offers perfect weak scaling:
The execution time for $\mathcal P \times N_s$ vectors on $\mathcal P$ processes is identical to the execution time for $N_s$ vectors on a single process. 
As a consequence, the pillar layout is always the most favorable vector layout for large $n$, when the cost of redistribution is amortized by the speedup of the Chebyshev filter.

\subsection{Memory requirements}
\label{sec:memory}

As explained earlier, for vectors layouts other than the stack layout the sparse matrix is duplicated in every process column,
such that $N_\mathrm{col}$ copies of the sparse matrix have to be stored in total. The memory requirements per process can be estimated as
\begin{equation}
 M = \frac{D}{\mathcal P} \Big( 3 N_s S_d  + N_\mathrm{col} \big( S_i + (S_i + S_d) n_\mathrm{nzr} \big)  \Big) \;.
\end{equation}
The first summand accounts for storage of the vector elements, where the factor $3$ arises from the three vectors used in the Chebyshev filter (see Alg.~\ref{alg:chebfilter}). The second summand accounts for storage of the matrix elements, say, in CSR format.
Redistribution of vectors can use the memory allocated for the Chebyshev filter, and has no additional memory requirements.

As a rough estimate, switching from the stack to the pillar layout with $N_\mathrm{col}=\mathcal P$ requires additional storage for the duplicated sparse matrices equivalent to $(\mathcal P -1) n_\mathrm{nzr}$ vectors per process.
These memory requirements become easily infeasible, which is why intermediate panel layouts with $N_\mathrm{col} < \mathcal P$ are relevant even though they do not achieve the same speedup as the pillar layout.

For the large-scale computations in the next section we use a matrix-free SpMV implementation, for which memory requirements are independent of the vector layout. Matrix-free implementations are readily available for stencil-like matrices~\cite{doi:10.1177/1094342020959423} such as the \texttt{Exciton} matrix, but can also be obtained for other typical quantum physics matrices such as the \texttt{Hubbard} matrix because of the tensor-product structure of the quantum many-particle Hilbert space~\cite{Da94,StronglyCorrelated}. Implementation details of matrix-free SpMV are outside of the scope of the present study. Nevertheless, the results of the present section show that implementation of a matrix-free SpMV is often worth the effort---not only because of node-level performance considerations, but to harness the speedup of vector layouts other than the standard stack layout.

\section{Filter diagonalization with two layers of parallelism}
\label{sec:FDlayers}

\begin{figure}
\hspace*{\fill}
\includegraphics[scale=0.9]{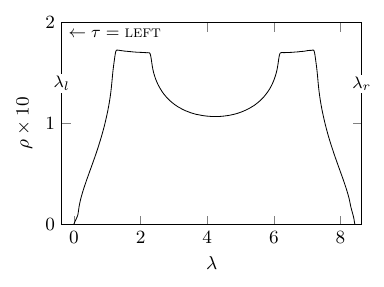}
\hspace*{\fill}
\includegraphics[scale=0.9]{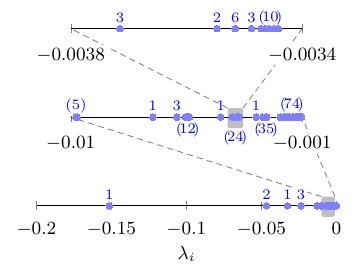}
\hspace*{\fill}
\caption{Left panel: Density of states $\rho(\lambda)$ of the \texttt{Exciton200} matrix ($D = 193\,443\,603$).
Right panel: Eigenvalues near the left end of the spectrum, for $\lambda_i < 0$, at three levels of magnification.
Blue numbers indicate the multiplicity of eigenvalues or, if in round brackets, the size of a cluster of nearly degenerate eigenvalues.
The width of the interval $[-0.0038,0.0034]$ on top is smaller than $5 \times 10^{-5}$ of the width of the entire spectrum; it still contains $24$ eigenvalues (counted with multiplicities).
}
\label{fig:Exciton}
\end{figure}

\begin{figure}
\hspace*{\fill}
\includegraphics[scale=0.9]{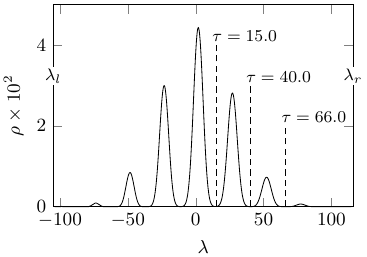}
\hspace*{\fill}
\includegraphics[scale=0.9]{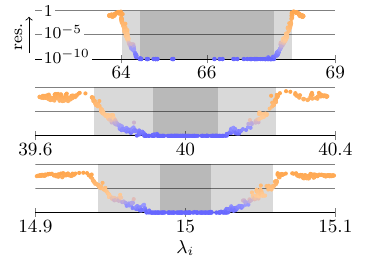}
\hspace*{\fill}
\caption{Left panel: Density of states $\rho(\lambda)$ of the \texttt{Hubbard16} matrix ($D = 165\,636\,900$).
Additional matrix parameters in \scamac{} notation are 
\texttt{U=25,ranpot=1} to create partially filled gaps with small density of states.
Right panel: Eigenvalues near the targets $\tau = 15.0, 40.0, 66.0$ inside of the spectrum. Distance to the horizontal axis indicates the residual of the eigenvectors, as shown in the top panel. Converged eigenvalues, with residual below $10^{-10}$, are found on the horizontal axis. The larger (smaller) gray rectangle indicates the search (target) interval.
}
\label{fig:Hubbard}
\end{figure}

The \texttt{Exciton} and \texttt{Hubbard} matrices in Table~\ref{tab:commcomp} are not merely artificial test cases for our communication metrics but appear in quantum physics research problems that require extraction of a large number of extremal or interior eigenvalues from tiny regions of the entire spectrum. In this respect, both problems pose a challenge for any eigensolver.
As typical examples of the matrices found in quantum physics applications, the
\texttt{Exciton} and \texttt{Hubbard} matrices have wildly different sparsity patterns and spectral properties but share a large matrix dimension.

In this section we use the \texttt{Exciton200} and \texttt{Hubbard16} matrix to show how typical large-scale eigenvalue computations are performed with FD.
After presentation of the general results in Sec.~\ref{sec:FD:results} we discuss the overall algorithmic performance in Sec.~\ref{sec:FD:performance} and formulate our key messages.
Sec.~\ref{sec:FD:speedup} is devoted to a more quantitative study of the speedup achieved for FD using the performance analysis of SpMV and the Chebyshev filter from the previous section.

\subsection{Computations and results}
\label{sec:FD:results}

The \texttt{Exciton} matrix appears in the modeling of an exciton, that is a bound electron-hole pair akin to a hydrogen atom
in a semiconductor, in the cuprous oxide $\text{Cu}_2\text{O}$. In this material, even-parity exciton states exhibit strong deviations from the hydrogen-like Rydberg series ``$E_n = - E_0/n^2$'' that can be traced back to spin-orbit coupling and crystal lattice effects on the kinetic energy of the electron and hole~\cite{Kl07,KFSSB14}. To compute the exciton states we thus cannot reuse the textbook solution of the hydrogen atom but have to solve the full lattice problem~\cite{ALF11,ExcitonII}, which leads to the \texttt{Exciton} matrices. The matrix parameter \texttt{L} controls the truncation of the infinite quantum mechanical state space to a finite lattice.
The matrix dimension is $D = 3 (2\,\text{\texttt{L}}+1)^3$. Computation of high excitonic states requires large values of \texttt{L}, here $\text{\texttt{L}}=200$.

The left panel in Fig.~\ref{fig:Exciton} shows the density of states (DOS) $\rho(\lambda)$ of the \texttt{Exciton} matrix,
computed with the kernel polynomial method (KPM)~\cite{WWAF06}.
The DOS gives the number of eigenvalues in an interval as $|\{\lambda_i \text{ with } a \le \lambda_i \le b\}| = D \int_a^b \rho(\lambda) \, d\lambda$.
The matrix dimension $D$ appears because we normalize the DOS to $\int_{-\infty}^\infty \rho(\lambda) \, d\lambda = 1$.
The exciton states of interest appear at the left (lower) end of the spectrum, for $\lambda_i < 0$, below the much broader spectrum of unbound electron-hole pairs that cannot be excluded from the lattice problem.
The values of $\lambda_i$ correspond to the physical exciton energies measured in electron volt.

The right panel in Fig.~\ref{fig:Exciton} shows the exciton states with eigenvalues $\lambda_i < 0$ as computed with FD.
The exciton states are characterized by multiple clusters of (nearly) degenerate eigenvalues, which move closer together as $\lambda_i$ approaches zero from below. This is reminiscent of the Rydberg series of the hydrogen atom. Resolution of eigenvalues within the clusters is required for comparison with optical measurements of the exciton emission spectra, which achieve high accuracy and thus require equal accuracy of the numerical computation~\cite{KFSSB14}.
Using many search vectors, FD is oblivious to the clustered structure of the spectrum and can fully resolve all eigenvalues.

The \texttt{Hubbard} matrix appears in many-body quantum physics, where the Hubbard model is a paradigmatic model of strongly correlated  electrons where Coulomb interaction leads to metal-insulator transitions or superconductivity~\cite{Da94,StronglyCorrelated}.
Recent research on, e.g., ultrafast control of quantum materials or many-body thermalization addresses physics associated with interior eigenvalues of the \texttt{Hubbard} (or related) matrices~\cite{doi:10.1126/science.aaa7432,RevModPhys.93.041002,10.1038/s41467-021-24726-0}.
It is characteristic of many-body quantum physics that the dimension $D = \binom{\text{\texttt{n\_sites}}}{\text{\texttt{n\_fermions}}}^2$
of the \texttt{Hubbard} matrix grows quickly with the physical problem size,
given by the matrix parameters \texttt{n\_sites}, \texttt{n\_fermions}.

The left panel in Fig.~\ref{fig:Hubbard} shows the DOS of the \texttt{Hubbard} matrix computed with KPM, the right panel shows the eigenvalues computed with FD.
To let the FD computations finish in reasonable time we chose (perhaps unphysical) matrix parameters (\texttt{U=25,ranpot=1} in \scamac{} notation; both measured in units of the hopping matrix element $t$), where partially filled gaps with small DOS appear in the spectrum.
Note that in physics terms we deal with the many-body DOS of the Hubbard model; for the \texttt{Hubbard16} matrix with $8$ electrons per spin orientation the many-body DOS for large $\texttt{U}$ has a peak in the center of the spectrum near $\lambda=0$ that corresponds to states with $8/2=4$ doubly occupied sites.
Computationally, in regions of large DOS the widths of the target and search interval become so tiny that the degree of the filter polynomial gets too large for our present demonstrations. The plain FD algorithm~\ref{alg:peigfex} has no strategy to remedy this situation; algorithmic improvements beyond the scope of the present work would become necessary. Therefore, we here use FD for targets $\tau$ in the partially filled gaps.

\begin{table}
\caption{Summary of FD with the \texttt{Exciton} (\texttt{Hubbard}) matrix from Figs.~\ref{fig:Exciton},~\ref{fig:Hubbard}, with $N_s=384$ ($N_s=512$) search vectors on $\mathcal P=64$ compute nodes, and $N_t=100$ requested target vectors.}
\label{tab:FD}
\medskip
\hspace*{\fill}
\begin{tabular}{p{1mm}@{\hspace{3mm}}ccrccrcc}
\toprule
& & & \multicolumn{1}{c}{no. of} & \multicolumn{2}{c}{relative width of} &  \multicolumn{1}{c}{conv.} & & \multicolumn{1}{c}{no.\,of\,{\big/}time} \\[-1pt]
& \smash{\raisebox{1.5ex}{matrix}} & \smash{\raisebox{1.5ex}{target}} & \multicolumn{1}{c}{SpMVs} & target int. & search int. &  \multicolumn{1}{c}{vecs.} & \smash{\raisebox{1.5ex}{runtime}} & \multicolumn{1}{c}{redistribution}\\
\midrule
\footnotesize 1: & \rule{0pt}{3ex}\texttt{Exciton200} & \textsc{left} & $16\,979$ & $4.0 \times 10^{-2}$ & $4.1 \times 10^{-2}$ & $138$ & 08:20:53 & 42\big/$3.9\%$ \\
\footnotesize 2: & \multicolumn{2}{c}{[with $N_t = 50$:} & $9\,939$ & $4.0 \times 10^{-2}$ & $4.1 \times 10^{-2}$ & 58 & 04:53:17 & 42\big/$6.4\%$\rlap{\,]}
\\[1ex]
\footnotesize 3: & \texttt{Hubbard16} & $\tau=15.0$ & $46\,347$ & $1.5 \times 10^{-4}$ & $5.3 \times 10^{-4}$ & 143 &  26:17:20 &  46\big/$0.6\%$  \\
\footnotesize 4: & \rule{5ex}{0.5pt} & $\tau=40.0$ & $12\, 984$ & $7.8 \times 10^{-4}$  & $2.2 \times 10^{-3}$ & 123 & 07:21:21 & 44\big/$2.1\%$ \\
\footnotesize 5: & \rule{5ex}{0.5pt} & $\tau=66.0$ & $4\,090$ & $1.4 \times 10^{-2}$ & $1.8 \times 10^{-2}$ & $101$ & 02:18:52 & 44\big/$7.2\%$ \\[0.5ex]
\bottomrule
\end{tabular}
\hspace*{\fill}
\end{table}

\subsection{Performance}
\label{sec:FD:performance}

Table~\ref{tab:FD} documents the algorithmic performance of FD for the computations in Figs.~\ref{fig:Exciton},~\ref{fig:Hubbard}.
The experiments were conducted as follows:
First, we implemented a matrix-free SpMV for the \texttt{Exciton} and \texttt{Hubbard} matrices.
Then, since memory is required only for storage of vectors,
we can use the pillar layout with $N_s/\mathcal P = 6$ ($8$) search vectors per compute node for the \texttt{Exciton} (\texttt{Hubbard}) matrix\footnote{On the Meggie cluster, each node has \SI{64}{\gibi\byte} main memory. The \texttt{Exciton} matrix has complex-valued entries (i.e., $S_d=16$) while the \texttt{Hubbard} matrix is real (i.e., $S_d=8$). Three copies of each search vector are required in the Chebyshev iteration in Alg.~\ref{alg:chebfilter}, resulting in \SI{51.9}{\gibi\byte} (\SI{29.6}{\gibi\byte}) memory consumption per node for the \texttt{Exciton200} (\texttt{Hubbard16}) matrix with $6$ ($8$) search vectors per node.
}.
In the pillar layout, the matrix-free implementation takes \SI{1.77}{\second} per SpMV for the \texttt{Exciton200} matrix and 
\SI{2.04}{\second} per SpMV for the \texttt{Hubbard16} matrix.
These execution times are consistent with values extrapolated for $N_p \to 1$ from the data in Fig.~\ref{fig:bench} (right panel), where we had to use $N_p \ge 4$ compute nodes due to memory demands arising from explicit storage of the sparse matrix.
We did not invest further effort into optimization of the matrix-free SpMV routines because already the unoptimized implementation suffices to demonstrate the importance of (avoiding) communication.

Note that the data in Table~\ref{tab:FD} support the two key assumptions made in the present work for the FD algorithm:
First, the degree of the filter polynomial is large since the width of the target and search interval is small.
FD requires about $20$ iterations for convergence, with about $2 \times 20 = 40$ redistributions of vectors (see last column).
In the course of the iterations, the target and search intervals shrink to their final sizes,
and the degree of the filter polynomial increases.
For the \texttt{Exciton200} matrix, the degree increases from $n=20$ in the first iteration to $n=1588$ in the final (twenty-first) iteration.
Second, as a consequence of the large degree of the filter polynomial, the total number of SpMVs is high and the cost of redistribution negligible (last column).
For the \texttt{Hubbard16} matrix, the last column shows how the percentage of the runtime spent on redistribution decreases even below $1\%$ 
as the degree of the filter polynomial increases for targets $\tau$ closer towards the center of the spectrum.

Table~\ref{tab:FD} also supports our two key messages for the practical application.
First, filter diagonalization works:
Despite the complicated structure of the spectrum (for the \texttt{Exciton} matrix) or the tiny size of the target interval (for the \texttt{Hubbard} matrix), FD successfully and reliably computes many eigenvectors and eigenvalues.
While we cannot claim that the plain FD scheme of Alg.~\ref{alg:peigfex} is particularly sophisticated, advanced, or guarantees fast convergence, it seems that a massive block algorithm such as FD may have its own merits even in large-scale eigenvalue computations.
Second, communication does not prevent scalability or parallel efficiency of large-scale eigenvalue computations if, but only if, one exploits the two layers of parallelism inherent to such computations.
This is probably the central message of the present work.

To see that this message is indeed true, note that the negligible cost of redistribution documented in Table~\ref{tab:FD} implies that the scalability of FD depends only on the scalability of the Chebyshev filter. As explained in Sec.~\ref{sec:amortize} the Chebyshev filter in the pillar layout shows perfect weak scaling, and for sufficiently many vectors per process column even strong scaling.
We can thus conclude that, owing to the pillar layout, the entire FD computation scales (almost) perfectly.

\subsection{Speedup}
\label{sec:FD:speedup}

\begin{figure}
\hspace*{\fill}%
\raisebox{-0.5\height}{\includegraphics[scale=1]{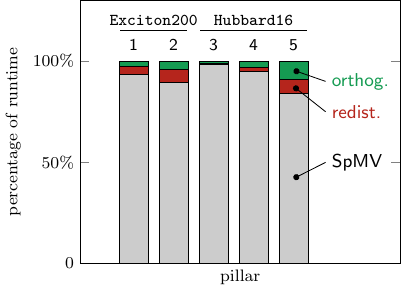}}%
\hspace*{\fill}%
\raisebox{-0.5\height}{\includegraphics[scale=1]{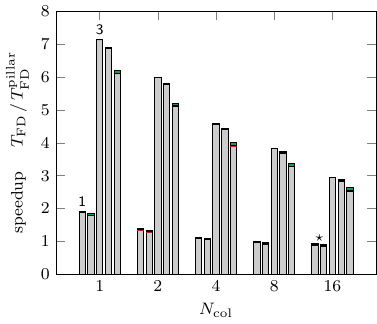}}%
\hspace*{\fill}%
\caption{Left panel: Relative contribution of SpMV, redistribution of vectors and orthogonalization to the total runtime of FD for the five application cases in Table~\ref{tab:FD}, following the enumeration $1$-$5$ in the table.
Right panel: Predicted runtime of FD in different vector layouts. For each case, the computation in the pillar layout ($N_\mathrm{col}=64$ as in Table~\ref{tab:FD}) is taken as the baseline, such that the height of each bar gives the speedup in the pillar layout relative to the given vector layout. The different contributions to the runtime are indicated with the colors from the left panel; the markings $1, 3, \star$ refer to the discussion in the text.}
\label{fig:speedup}
\end{figure}

Our analysis from Sec.~\ref{sec:layers} shows how to understand the performance of the Chebyshev filter through a combination of empirical benchmark data and a priori metrics for the basic operations SpMV and redistribution of vectors.
We can apply the same principle to the entire FD algorithm with total runtime
\begin{equation}\label{RuntimeFD}
 T_\mathrm{FD} = T_\mathrm{init} + n_\mathrm{SpMV} T_\mathrm{SpMV} + n_\mathrm{iter} (T_\mathrm{ortho} + 2 T_\mathrm{rd}) \;.
\end{equation}
In this expression, $n_\mathrm{iter}$ is the number of outer iterations of the FD algorithm (corresponding to the repeat loop in Alg.~\ref{alg:peigfex}, usually with $n_\mathrm{iter} \approx 20$),
and $n_\mathrm{SpMV}$ the total number of SpMVs in the Chebyshev filter accumulated over all iterations.

The execution times $T_\mathrm{SpMV}$ of SpMV and $T_\mathrm{rd}$ for redistribution of vectors have been measured and analyzed in Sec.~\ref{sec:layers}; both depend on the vector layout.
The execution time $T_\mathrm{ortho}$ of orthogonalization has been measured in the FD runs in the present section; it does not depend on the vector layout.
The time $T_\mathrm{init}$ for setup and initialization, which includes generation of the matrix (or of the matrix parameters for the matrix-free SpMV) via \scamac{}, is always negligible. 

For the computations reported in Table~\ref{tab:FD} we consistently find $T_\mathrm{ortho} \approx \SI{40}{\s}$ (without having applied any optimizations), that is,
orthogonalization is slow compared to SpMV ($T_\mathrm{ortho}/T_\mathrm{SpMV} \approx 20$).
Nevertheless, the contribution of orthogonalization to the total runtime of FD remains small (here, below $10\%$)
since orthogonalization is executed only once per outer iteration, that is only once every $n_\mathrm{SpMV}/n_\mathrm{iter}$ SpMVs. All FD computations are in the regime $n_\mathrm{SpMV} T_\mathrm{SpMV} \gg n_\mathrm{iter} T_\mathrm{ortho}$.
Note that in this respect FD resembles other communication-avoiding techniques that execute multiple SpMVs before a potentially slower operation such as orthogonalization~\cite{10.1145/1654059.1654096,10.1007/978-3-319-17353-5_2,ACER201671,8821043}. For FD, this procedure is inherently built into the algorithm.

The relevance of the different contribution to the total runtime in Eq.~\eqref{RuntimeFD} is shown in the left panel of Fig.~\ref{fig:speedup}.
Even for the shorter runs (cases $2, 5$ in the figure and in Table~\ref{tab:FD}) the runtime is dominated by SpMV.
We here observe, once more, the central characteristics of FD for large-scale eigenvalue computations:
(i) FD is about the Chebyshev filter, (ii) the Chebyshev filter is about SpMV, (iii) SpMV is about communication.
All large-scale FD computations are in a regime where the speedup of the entire FD computation is mainly determined by the speedup $S$ of the Chebyshev filter, which we have analyzed in Sec.~\ref{sec:amortize}.

In fact, we have assembled all necessary data to quantify the speedup of FD in the pillar layout,
without having to repeat either the benchmarks from Sec.~\ref{sec:layers} or the full FD computations.
The predicted FD runtime in the different vector layouts according to Eq.~\eqref{RuntimeFD} and the earlier benchmarks is shown in the right panel of Fig.~\ref{fig:speedup}.
This figure essentially reflects the values from Table~\ref{tab:amortize}, with small corrections due to the additional orthogonalization in each outer iteration.

The bars marked with $1$, $3$ in Fig.~\ref{fig:speedup} correspond to a computation in the stack layout.
The best possible speedup in the pillar layout according to Table~\ref{tab:amortize} is $s=2.02$ for the \texttt{Exciton200} and $s=7.25$ for the \texttt{Hubbard16} matrix. The actual speedup is slightly smaller due to the necessary redistribution of vectors, and can be deduced from Eq.~\eqref{BigS}. For example for the first case, with $n_\mathrm{SpMV}/n_\mathrm{iter} = 16979/21 \approx 809$ (see first row in Table~\ref{tab:FD}), Eq.~\eqref{BigS} gives the speedup $S \approx 1.88$. This is the value observed in Fig.~\ref{fig:speedup}. For the second case, the speedup is close to the asymptotic value $S \approx s = 7.25$. In all cases, the runtime of orthogonalization can be neglected.

For the two bars marked with a $\star$ symbol the speedup relative to the stack layout is even a bit larger than that achieved with the pillar layout
(we have $s=2.13$ for $N_\mathrm{col}=16$ but $s=2.02$ for $N_\mathrm{col}=64$ in Table~\ref{tab:amortize}).
This reflects the superior node-level performance of the optimized SpMV of the \ghost{} library over that of the matrix-free SpMV used in the pillar layout. In these cases, where the communication overhead of SpMV is reduced by using a general panel layout, optimization of the node-level performance of SpMV in the \ghost{} library pays off. Remarkably, even for the \texttt{Exciton} matrices the performance of an unoptimized SpMV in the pillar layout is significantly better than that of an optimized SpMV in the stack layout.

We can summarize Fig.~\ref{fig:speedup} as follows:
The number of SpMVs in the present FD computations is two to four orders of magnitude larger than the break-even point $n^*$ reported in Table~\ref{tab:amortize}, such that we almost achieve the asymptotic speedup $S \simeq s$ of the Chebyshev filter.
In short, for the \texttt{Exciton200} (\texttt{Hubbard16}) matrix the FD computation with two layers of parallelism is twice (seven times) as fast as a computation that uses only the one layer of parallelism offered by the standard stack layout.

Finally note that a significant speedup is already achieved with our default SpMV implementation.
Careful optimization of the matrix-free SpMV routines could boost the node-level performance, and thus increase the speedup $S$.
Higher node-level performance with faster local operations makes it even more important to avoid communication, which further strengthens our message.

\section{Summary \& Outlook}
\label{sec:summary}

The main obstacle to scalability of large-scale distributed computations is communication.
Preserving scalability and parallel efficiency thus requires strategies to minimize communication.
In the present paper we introduce such a strategy for large-scale eigenvalue computations, using filter diagonalization (FD) as an example of an eigensolver with many search vectors. In such eigensolvers we can identify two orthogonal layers of parallelism, which describe how vectors and sparse matrices are distributed in a general panel layout.

To minimize communication we must compare the different distributed vector layouts for the central operation of the eigensolver, sparse matrix-vector multiplication (SpMV). 
Communication is maximal in the standard stack layout, but can completely be avoided in the opposite extreme of the pillar layout.
However, it is not obvious that the stack layout should be replaced by the pillar layout since SpMV and the second major operation in the eigensolver, which is orthogonalization, have competing communication patterns. Therefore, the question arises whether the cost of redistribution of vectors between different layouts is amortized by the speedup of SpMV.

To answer this question we introduce a metric $\chi$ to quantify the communication volume of SpMV, which is directly derived from the matrix sparsity pattern. Benchmarks show that our model for the communication overhead of SpMV, based on the metric $\chi$, can accurately predict the speedup of SpMV in the different vector layouts. We can thus identify the conditions under which it is favorable to give up the standard stack layout in favor of a panel (or pillar) layout.
Quite generally, these conditions are fulfilled for matrices with a ``wide'' sparsity pattern and large values of $\chi$. But even for matrices with a ``narrow'' sparsity pattern and small values of $\chi$, a speedup is achieved if a large number of SpMVs is executed.

The general scenario developed here is clearly observed in the exemplary FD computations for the large-scale \texttt{Exciton} and \texttt{Hubbard} matrices.
Even for a stencil-like \texttt{Exciton} matrix, with values $\chi < 1$, the communication overhead of SpMV in the stack layout is large enough to achieve a speedup of $2$ by switching to the pillar layout.
For matrices of \texttt{Hubbard}-type, with $\chi > 1$, we can achieve almost an order of magnitude speedup.
Note that for very large matrix dimension it will be impossible to use the pillar layout because of memory requirements, even with a matrix-free SpMV.
In essence, the pillar layout trades memory for performance; if the memory requirements cannot be fulfilled, a general panel layout with reduced number of process columns must be chosen.
These layouts are less favorable than the pillar layout but still improve on the standard stack layout, where only a single layer of parallelism is considered.

For the present paper we chose FD as our algorithmic example of an eigensolver with two layers of parallelism.
Since the algorithmic performance of FD is not as good as one might expect from a general-purpose eigensolver,
it is important to ask whether the current results on computational performance, in terms of preserving or restoring scalability and parallel efficiency, generalize to other algorithms. The two assumptions underlying our results are 
(i) we deal with a block eigensolver with multiple or many search vectors, and
(ii) multiple SpMVs are executed in a row (more than the break-even point $n^*$ derived in Eq.~\eqref{Favorable}).
Under these assumptions, our results generalize to different eigensolvers or similar algorithms for other sparse liner algebra problems.
One class of algorithms that might improve on the algorithmic shortcomings of FD are hybrid algorithms that combine Chebyshev filters and a few Arnoldi steps to construct a Krylov subspace. Whether such algorithms can be successfully applied to relevant large-scale eigenvalue problems, for example to the Hubbard model in regions with large DOS, remains to be demonstrated.

Although driven by the constraints of large-scale applications, where communication in SpMV is a major obstacle, the focus of the present paper rather lies on the theoretical analysis and empirical benchmarks necessary to establish the communication metrics and vector layouts.
Immediately, the question arises how future work might achieve integration of the ideas developed here into production-level (eigensolver) code.
We cannot provide a full answer to this question, but we will draw a few conclusions from the present work in the following.

First, there is the issue of matrix-based versus matrix-free SpMV. Here, we use a matrix-free SpMV in the pillar layout
in order to save memory.
In the pillar layout, SpMV does not require communication, which simplifies implementation of the matrix-free SpMV. For general applications and general panel layouts, the communication metrics of the present work allow us to predict the potential speedup prior to benchmarks or large-scale computations, and thus to decide whether the implementation effort of a matrix-free SpMV will be justified. To reiterate a central observation: Even for matrices with ``narrow'' sparsity patterns, for which at first sight communication in SpMV  does not seem to be of importance, a substantial speedup can be achieved.

Second, one might ask whether other techniques such as reordering of the matrix via graph partitioning~\cite{Karypis2011,ZoltanGraph} solve the communication problem of distributed SpMV. To date, our attempts in this direction have been consistently unsuccessful.
For matrices of \texttt{Exciton}-type, the minimal communication volume is that of a regular 3D stencil and not reduced further by reordering or partitioning.
For matrices of \texttt{Hubbard}-type, and equally so for many other matrices from many-body quantum mechanics, the sparsity pattern corresponds to a graph with high effective dimension, far exceeding the geometric dimension of the problem under consideration. Graph partitioning does not seem to be effective for such matrices.

A related aspect is the balancing of communication volume by adjusting the distribution of rows onto processes. 
Also this task can be addressed by inspection of the sparsity pattern and the communication metrics: Recall that a communication imbalance is detected by the discrepancy of the three metrics $\chi_{1,2,3}$. Implementation of a practical algorithm that allows for balancing of the communication volume also for truly large matrix dimensions, on the other hand, is far from trivial.

In conclusion, we present a theoretical analysis of communication in SpMV that is based entirely on the matrix structure encoded in the sparsity pattern, and a practical approach to reduce communication that is entirely general and does not depend on the particular structure of the matrix.
In practice,
the following guidelines apply: If the matrix is ``well behaved'' with small communication metrics, communication does not harm the performance of SpMV. If, on the other hand, the communication metrics are large, a different vector layout is required in order to achieve the desired speedup.

\section{Code availability}

All data in the present work is generated with the FD code at \url{https://www.bitbucket.org/alvbit/twolayerfd}.
The code is built on top of the \ghost{} library (\url{https://www.bitbucket.org/essex/ghost}) for distributed sparse-matrix and vector operations~\cite{GHOST},
and uses \scamac{} (\url{https://www.bitbucket.org/essex/matrixcollection}) for the generation of the different sparse matrices.
The repository contains the \texttt{scamac\_count\_commvol} tool to compute the communication metrics and volume directly from the sparsity pattern.

\begin{acks}
The authors would like to thank Dominik Ernst for help with the \ghost{} library.
This project was funded by the \textit{Competence Network for Scientific High Performance Computing in Bavaria} (KONWIHR).
The authors gratefully acknowledge the scientific support and HPC resources provided by the Erlangen National High Performance Computing Center (NHR@FAU) of the Friedrich-Alexander-Universität Erlangen-Nürnberg (FAU). The hardware is funded by the German Research Foundation (DFG).
\end{acks}

\appendix

\section{Additional benchmark data}
\label{app:scamac}

\begin{figure}
\hspace*{\fill}
\includegraphics[width=\textwidth]{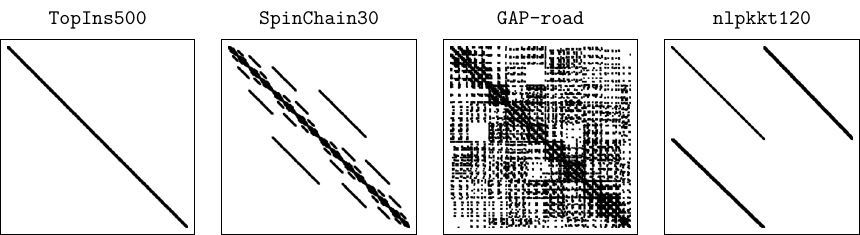}
\hspace*{\fill}
\caption{From left to right, sparsity pattern of the \texttt{TopIns500} and \texttt{SpinChain30} matrix from \scamac{}~\cite{SCAMAC} (see Table~\ref{tab:various} and Fig.~\ref{fig:bench_various}), and of the \texttt{GAP-road} and \texttt{nlpkkt120} matrix from the SuiteSparse collection~\cite{SuiteSparse} (see Table~\ref{tab:various_irreg} and Fig.~\ref{fig:bench_irregular}).
}
\label{fig:sparsity2}
\end{figure}

In the appendix we want to extend the results for our communication model from the main text with benchmarks for a few additional matrices.
We use four matrices with regular sparsity patterns, similar to the \texttt{Exciton} and \texttt{Hubbard} matrices in the main text, which are typical for matrices from (quantum) physics applications, 
and two matrices with irregular sparsity patterns, as they occur in other application domains.
Selected sparsity patterns are shown in Fig.~\ref{fig:sparsity2}.

\subsection{Regular sparsity patterns} 

\begin{table}
\caption{Communication metrics for additional \scamac{} matrices.}
\label{tab:various}
\medskip
\hfill
\begin{tabular}{lrrrrr}
\toprule
matrix  \hspace{2cm} & $N_p$  & $\chi_{1,3}$ & $\chi_2$ \\ 
\midrule
\texttt{TopIns,Lx=100,}           & 2 & 0.02 & 0.02       \\
\texttt{\quad Ly=100,Lz=100}     &  4 & 0.08 & 0.06        \\
                                 &  8 & 0.16 & 0.14       \\
\quad $D = 4\,000\,000$         &  16 & 0.32 & 0.30     \\
\quad $n_\mathrm{nzr} = 11.88$   & 32 & 0.64 & 0.62     \\
                                 & 64 & 1.28 & 1.26 \\[4ex]
\texttt{TopIns,Lx=500,}           & 2 & 0.00 & 0.00     \\
\texttt{\quad Ly=500,Lz=500}     &  4 & 0.02 & 0.01   \\
                                 &  8 & 0.03 & 0.03   \\
\quad $D = 500\,000\,000$       &  16 & 0.06 & 0.06 \\
\quad $n_\mathrm{nzr} = 11.98$   & 32 & 0.13 & 0.12   \\
                                 & 64 & 0.26 & 0.25   \\
\bottomrule
\end{tabular}
\hfill{}
\begin{tabular}{lrrr}
\toprule
matrix \hspace{2cm}  & $N_p$  & $\chi_{1,3}$ & $\chi_2$ \\ 
\midrule
\texttt{SpinChainXXZ,}              &  2 & 0.52 & 0.52    \\
\texttt{\;n\_sites=24,n\_up=12}    &  4 & 1.50 & 1.01    \\
                                    &  8 & 2.51 & 1.52    \\
\quad $D = 2\,704\,156$             & 16 & 3.40 & 2.00   \\
\quad $n_\mathrm{nzr} = 13$         & 32 & 4.18 & 2.49   \\
                                    & 64 & 5.15 & 3.05  \\[4ex]
\texttt{SpinChainXXZ,}               & 2 & 0.52 & 0.52   \\
\texttt{\;n\_sites=30,n\_up=15}    &  4 & 1.50 & 1.01 \\
                                    &  8 & 2.49 & 1.51  \\
\quad $D = 155\,117\,520$          &  16 & 3.43 & 1.99 \\
\quad $n_\mathrm{nzr} = 16$         & 32 & 4.27 & 2.47   \\
                                    & 64 & 5.10 & 3.03  \\
\bottomrule
\end{tabular}
\hfill{}
\end{table}

Table~\ref{tab:various} shows the communication metrics for the \scamac{} matrices \texttt{TopIns} (which we used in Ref.~\cite{Pieper16})
and \texttt{SpinChainXXZ} (which we used in Ref.~\cite{zoellner2015}).
From the sparsity patterns in Fig.~\ref{fig:sparsity2} 
we see that the \texttt{TopIns} matrices resemble the \texttt{Exciton} matrices, with a stencil-like structure and ``narrow'' sparsity pattern that results in small communication volume.
The \texttt{SpinChain} matrices resemble the \texttt{Hubbard} matrices with a ``wide'' sparsity pattern and large communication volume.

\begin{figure}
\includegraphics[scale=0.8]{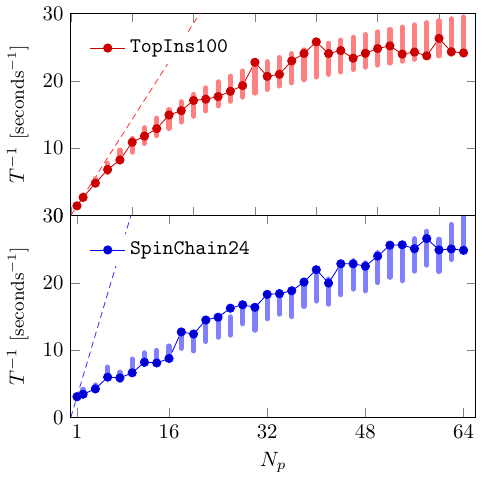}
\hspace*{\fill}
\includegraphics[scale=0.8]{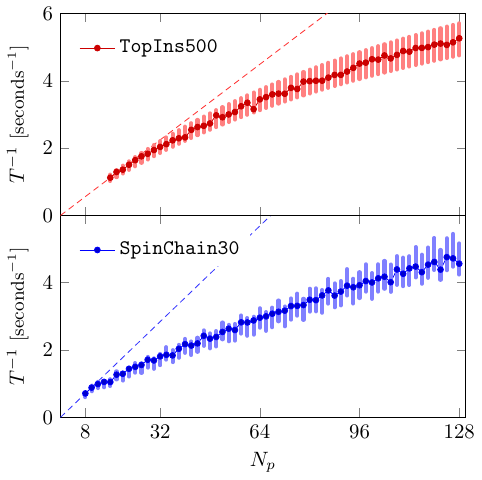}
\caption{Same plot as in Fig.~\ref{fig:bench}, now for the \texttt{SpinChainXXZ} and \texttt{TopIns} matrices from Tab.~\ref{tab:various}.
In the right panel, the execution time for $N_p=8$ (\texttt{SpinChain30}) or $N_p=16$ (\texttt{TopIns500}) processes is used as a baseline. The left (right) panel uses $n_b=64$ ($n_b=8$) vectors.
}
\label{fig:bench_various}
\end{figure}

The SpMV benchmarks and model predictions are shown in Fig.~\ref{fig:bench_various}.
Again, the theoretical data of Eq.~\eqref{EstimateT} agree well with the benchmark data, such that conclusions analogous to those in the main text can be drawn from the communication metrics and our communication model for SpMV.

We like to emphasize that further benchmark data or tables for the communication metric $\chi$ can be generated with the tools in the FD and \scamac{} repositories, and thus need not be included here.

\subsection{Irregular sparsity patterns} 
\label{app:irregular}

As examples for matrices with irregular sparsity patterns we choose the  \texttt{GAP-road} and \texttt{nlpkkt120} matrix from the SuiteSparse collection~\cite{SuiteSparse}. Other choices are possible and lead to comparable results. Note, however, that the present study requires larger matrices than those typical for the SuiteSparse collection. For small matrices, our assumption that communication is volume bound instead of latency bound is no longer valid.

\begin{table}
\caption{Communication metrics for two irregular matrices 
from the SuiteSparse collection~\cite{SuiteSparse}.} 
\label{tab:various_irreg}
\medskip
\hfill
\begin{tabular}{lrrrr}
\toprule
matrix \hspace{1.5cm}  & $N_p$  & $\chi_1$ & $\chi_2$ & $\chi_3$ \\ 
\midrule
\texttt{GAP-road}                &  2 & 0.04 & 0.03 & 0.04 \\      
                                 &  4 & 0.19 & 0.13 & 0.19 \\
                                 &  8 & 0.53 & 0.33 & 0.45 \\
\quad $D = 23\,947\,347$         & 16 & 2.36 & 0.70 & 1.24 \\
\quad $n_\mathrm{nzr} = 2.41$    & 32 & 6.32 & 1.09 & 1.78 \\
                                 & 64 & 7.60 & 1.11 & 1.87 \\                     
\bottomrule
\end{tabular}
\hfill{}
\begin{tabular}{lrrrr}
\toprule
matrix \hspace{1.5cm}  & $N_p$  & $\chi_1$ & $\chi_2$ & $\chi_3$ \\ 
\midrule
\texttt{nlpkkt120}               &  2 & 20.67 & 0.98 & 1.00 \\
                                 &  4 & $\infty$ & 1.02 & 1.07 \\
                                 &  8 & $\infty$ & 1.06 & 1.10 \\
\quad $D = 3\,542\,400$              & 16 & $\infty$ & 1.12 & 1.17 \\
\quad $n_\mathrm{nzr} = 26.85$   & 32 & $\infty$ & 1.26 & 1.31 \\
                                 & 64 & $\infty$ & 1.52 & 1.58 \\
\bottomrule
\end{tabular}
\hfill{}
\end{table}

Table~\ref{tab:various_irreg} shows the communication metrics for the \texttt{GAP-road} and \texttt{nlpkkt120} matrix.
Now the three communication metrics $\chi_{1,2,3}$ differ, in contrast to our discussion in Sec.~\ref{sec:metric}.
This indicates the irregularity of the sparsity pattern. It also indicates that for these matrices additional effort should be invested into graph partitioning 
or matrix reordering to avoid severe load or communication imbalances. For typical (quantum) physics matrices with $\chi_1 \approx \chi_2 \approx \chi_3$, like those that motivate the present study, these efforts are not immediately 
 required.

Although the discrepancy between the different $\chi$ metrics does not invalidate our model it warrants closer scrutiny.
First, we see from the large values of $\chi_1$ that for the present two matrices SpMV is clearly communication instead of memory bound.
For the \texttt{nlpkkt120} matrix, the $\chi_1 = \infty$ value is associated with the south east quadrant of the matrix. To evaluate SpMV for rows in the lower half of the matrix all vector entries have to be communicated, but no local vector entries are used---the worst-case scenario if communication should be avoided.
Second, we see that in spite of the irregular sparsity pattern, the amount of communication is not necessarily very large: The metrics $\chi_{2,3}$ are strictly smaller than for the \texttt{Hubbard} or \texttt{SpinChain} matrices.
Third, we see that still $\chi_2 \approx \chi_3$: The average and maximal communication volume per process remain comparable.
For comparison with Table~\ref{tab:commvol}, the communication volume for the matrices in the appendix is given in absolute numbers in Table~\ref{tab:commvol_appendix}.

\begin{table}
\caption{Average and maximal communication volume per process and SpMV for the matrices in the benchmarks in Figs.~\ref{fig:bench_various} and \ref{fig:bench_irregular}.} \label{tab:commvol_appendix}
\medskip
\hspace*{\fill}
\begin{tabular}{lr@{\hspace{5mm}}rr}
\toprule
 & & \multicolumn{2}{c}{communication volume} \\
\smash{\raisebox{1.5ex}{matrix}} \hspace{1cm}  & \smash{\raisebox{1.5ex}{$N_p$}}  & \multicolumn{1}{r}{average} & \multicolumn{1}{r}{maximal} \\ 
\midrule
\texttt{TopIns100}
 &  2 & \SI{39.1}{\mebi\byte} & \SI{39.1}{\mebi\byte} \\
 & 64 & \SI{76.9}{\mebi\byte} & \SI{78.1}{\mebi\byte} \\[0.5ex]
\texttt{SpinChain24}
 &  2 & \SI{344.4}{\mebi\byte} & \SI{344.4}{\mebi\byte} \\
 & 64 & \SI{62.8}{\mebi\byte} & \SI{106.3}{\mebi\byte} \\[0.5ex]
\texttt{TopIns500}
 & 16 & \SI{228.9}{\mebi\byte} & \SI{244.1}{\mebi\byte} \\
 & 64 & \SI{240.3}{\mebi\byte} & \SI{244.1}{\mebi\byte} \\[0.5ex]
\texttt{SpinChain30}
 &  8 & \SI{1.74}{\gibi\byte} & \SI{2.88}{\gibi\byte} \\
 & 64 & \SI{448.5}{\mebi\byte} & \SI{753.7}{\mebi\byte} \\
\texttt{GAP-road}
 &  2 & \SI{194.3}{\mebi\byte} & \SI{205.5}{\mebi\byte} \\
 & 64 & \SI{203.6}{\mebi\byte} & \SI{341.7}{\mebi\byte} \\[0.5ex]
\texttt{nlpkkt120}
 &  2 & \SI{851.0}{\mebi\byte} & \SI{864.8}{\mebi\byte} \\
 & 64 & \SI{41.2}{\mebi\byte} & \SI{42.6}{\mebi\byte} \\
\bottomrule
\end{tabular}
\hspace*{\fill}
\end{table}

\begin{figure}
\includegraphics[scale=0.8]{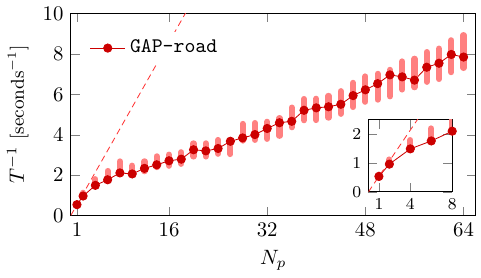}
\hspace*{\fill}
\includegraphics[scale=0.8]{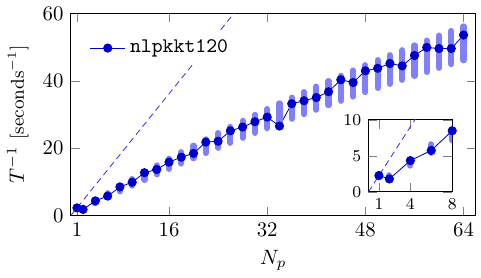}
\caption{Same plot as in Figs.~\ref{fig:bench},~\ref{fig:bench_various}, now for the \texttt{GAP-road} and \texttt{nlpkkt120} matrices from Tab.~\ref{tab:various_irreg}. Both benchmarks use $n_b=64$ vectors.
The insets show the results for $1 \le N_p \le 8$ in magnification.
}
\label{fig:bench_irregular}
\end{figure}

If we go back to the communication model we see that we should evaluate Eq.~\eqref{EstimateT} with the metric $\chi_3$ when we ask for the total SpMV runtime.
The SpMV benchmarks and model predictions are shown in Fig.~\ref{fig:bench_irregular}.
Again, and in spite of the irregular sparsity patterns and differing values of the communication metrics, the theoretical values of Eq.~\eqref{EstimateT} agree well with the benchmark data. In fact, by construction of our model, this can be expected for any situation with significant communication in the SpMV.

Two additional observations can be made in Fig.~\ref{fig:bench_irregular}.
For the \texttt{GAP-road} matrix we see that $\chi_3$ remains small for $N_p=2$ and becomes large only for $N_p \ge 4$.
Recall that ``small'' and ``large'' requires comparison of $\chi_3$ with the ratio $b_c/b_m$, see for example Eq.~\eqref{PiParallel}.
Indeed, the inset shows that SpMV scales well from $N_p=1$ to $N_p=2$, but ceases to scale for $N_p \ge 4$ processes.

For the \texttt{nlpkkt120} matrix, SpMV scales very well for $N_p \ge 2$. What we observe here is that $\chi_3 \approx 1$ remains large but almost constant for $2 \le N_p \le 64$. 
Then, according to Eq.~\eqref{EstimateT}, $T \propto N_p^{-1}$, but the large value of $\chi_3$ tells us that the excellent scaling for $N_p \ge 2$ corresponds to poor performance:
Prior to any benchmarks Eq.~\eqref{PiParallel} allows us to estimate that the parallel efficiency $\Pi \lesssim b_c/b_m$ for $\chi \approx 1$ (in our computations $b_c/b_m < 0.1$, see Table~\ref{tab:model_various}).
Indeed, the inset shows that from $N_p=1$ to $N_p=2$, that is from a situation without communication ($\chi = 0$) to a situation with much communication ($\chi_3 \approx 1$), the runtime of SpMV even increases. 

Of course, the two observations made here once more support our message for the practical application: If $\chi$ is large try to keep SpMV local, perhaps using the strategies from the main text.

\subsection{Model parameters}

We conclude the appendix with a comparison of the values of the model parameters $b_m$, $\kappa$, $b_c$ listed in Table~\ref{tab:model_various}. They should be compared also with the parameters in Table~\ref{tab:model} in the main text.
All model parameters assume reasonable and consistent values across the different matrices.
Just as for the benchmarks in the main text, the dependence of the runtime $T$ on the number of processes $N_p$ in Figs.~\ref{fig:bench_various},~\ref{fig:bench_irregular} is connected with the variation of the communication metric $\chi$, computed prior to the benchmarks without a fit to the benchmark data.
In short, our model applies.

\begin{table}
\caption{Model parameters $b_m$, $\kappa$, $b_c$ for the data in Figs.~\ref{fig:bench_various},~\ref{fig:bench_irregular},  with the same description as in Table~\ref{tab:model}.}
\label{tab:model_various}
\medskip
\hspace*{\fill}
\begin{tabular}{lccc}
\toprule
\multicolumn{1}{c}{matrix} &\rule{1ex}{0pt} $b_m$ [\si{\giga\byte\per\second}] \rule{1ex}{0pt} & $\kappa$ & $b_c$ [\si{\giga\byte\per\second}] \\
\midrule
\texttt{TopIns100} &  53.3 & 8.28   & 3.10 \\
\texttt{TopIns500} & $*$ & $*$ &  2.82 \\[0.5ex]
\texttt{SpinChain24} & $*$ & 12.2 & 3.52 \\
\texttt{SpinChain30} & $*$ & $*$ & 2.68 \\[0.5ex]
\texttt{GAP-road} & $*$ & 8.00 & 3.80 \\
\texttt{nlpkkt120} & $*$ & 12.3 & 3.52 \\
\bottomrule
\end{tabular}
\hspace*{\fill}
\end{table}



\end{document}